\input harvmac

\def\hat{\widehat}
\def\CF{{\cal F}}
%
\let\includefigures=\iftrue
%
%
%
\newfam\black
\input rotate
\input epsf
\noblackbox
%
%
\includefigures
\message{If you do not have epsf.tex (to include figures),}
\message{change the option at the top of the tex file.}
\def\figin{\epsfcheck\figin}\def\figins{\epsfcheck\figins}
\def\epsfcheck{\ifx\epsfbox\UnDeFiNeD
\message{(NO epsf.tex, FIGURES WILL BE IGNORED)}
\gdef\figin##1{\vskip2in}\gdef\figins##1{\hskip.5in}
\else\message{(FIGURES WILL BE INCLUDED)}%
\gdef\figin##1{##1}\gdef\figins##1{##1}\fi}
\def\DefWarn#1{}
\def\N{{\cal N}}
\def\figinsert{\goodbreak\midinsert}
\def\ifig#1#2#3{\DefWarn#1\xdef#1{fig.~\the\figno}
\writedef{#1\leftbracket fig.\noexpand~\the\figno}%
\figinsert\figin{\centerline{#3}}\medskip\centerline{\vbox{\baselineskip12pt
\advance\hsize by -1truein\noindent\footnotefont{\bf
Fig.~\the\figno:} #2}}
\bigskip\endinsert\global\advance\figno by1}
\else
\def\ifig#1#2#3{\xdef#1{fig.~\the\figno}
\writedef{#1\leftbracket fig.\noexpand~\the\figno}%
\global\advance\figno by1} \fi
\def\yboxit#1#2{\vbox{\hrule height #1 \hbox{\vrule width #1
\vbox{#2}\vrule width #1 }\hrule height #1 }}
\def\fillbox#1{\hbox to #1{\vbox to #1{\vfil}\hfil}}
\def\ybox{{\lower 1.3pt \yboxit{0.4pt}{\fillbox{8pt}}\hskip-0.2pt}}

\def\rightarrowbox#1#2{
  \setbox1=\hbox{\kern#1{${ #2}$}\kern#1}
  \,\vbox{\offinterlineskip\hbox to\wd1{\hfil\copy1\hfil}
    \kern 3pt\hbox to\wd1{\rightarrowfill}}}

\def\p{\partial}

\def\half{{1\over 2}}
\def\Tr{{{\rm Tr~ }}}
\def\tr{{\rm tr\ }}

\def\CF{{\cal F}}

\def\CM{{\cal M}}
\def\CN{{\cal N}}
\def\CO{{\cal O}}

\def\CP{{\cal P}}

\def\tilde{\widetilde}

\def\II{\relax{I\kern-.10em I}}

\def\bar{\overline}

\def\IZ{\relax\ifmmode\mathchoice
{\hbox{\cmss Z\kern-.4em Z}}{\hbox{\cmss Z\kern-.4em Z}}
{\lower.9pt\hbox{\cmsss Z\kern-.4em Z}} {\lower1.2pt\hbox{\cmsss
Z\kern-.4em Z}}\else{\cmss Z\kern-.4em Z}\fi}
\def\IB{\relax{\rm I\kern-.18em B}}
\def\IC{{\relax\hbox{$\inbar\kern-.3em{\rm C}$}}}
\def\ID{\relax{\rm I\kern-.18em D}}
\def\IE{\relax{\rm I\kern-.18em E}}
\def\IF{\relax{\rm I\kern-.18em F}}
\def\IG{\relax\hbox{$\inbar\kern-.3em{\rm G}$}}
\def\IGa{\relax\hbox{${\rm I}\kern-.18em\Gamma$}}
\def\IH{\relax{\rm I\kern-.18em H}}
\def\II{\relax{\rm I\kern-.18em I}}
\def\IK{\relax{\rm I\kern-.18em K}}
\def\IN{\relax{\rm I\kern-.18em N}}
\def\IP{\relax{\rm I\kern-.18em P}}

%
\def\inbar{\,\vrule height1.5ex width.4pt depth0pt}

\def\p{\partial}

\font\cmss=cmss10 \font\cmsss=cmss10 at 7pt
\def\IR{\relax{\rm I\kern-.18em R}}

\def\lp10{l_P^{10}}
\def\lp11{l_P^{11}}
\def\R11{R_{11}}

%
%
\lref\VenezianoAH{ G.~Veneziano and S.~Yankielowicz, ``An
Effective Lagrangian For The Pure N=1 Supersymmetric Yang-Mills
Theory,'' Phys.\ Lett.\ B {\bf 113}, 231 (1982).
}

\lref\GopakumarKI{ R.~Gopakumar and C.~Vafa, ``On the gauge
theory/geometry correspondence,'' Adv.\ Theor.\ Math.\ Phys.\ {\bf
3}, 1415 (1999) [arXiv:hep-th/9811131].
}

\lref\VafaWI{ C.~Vafa, ``Superstrings and topological strings at
large N,'' J.\ Math.\ Phys.\  {\bf 42}, 2798 (2001)
[arXiv:hep-th/0008142].
}

\lref\DiFrancescoNW{ P.~Di Francesco, P.~Ginsparg and
J.~Zinn-Justin, ``2-D Gravity and random matrices,'' Phys.\ Rept.\
{\bf 254}, 1 (1995) [arXiv:hep-th/9306153].
}

\lref\SeibergJQ{ N.~Seiberg, ``Adding fundamental matter to
'Chiral rings and anomalies in  supersymmetric gauge theory',''
arXiv:hep-th/0212225.
}

\lref\HofmanBI{ C.~Hofman, ``Super Yang-Mills with flavors from
large N(f) matrix models,'' arXiv:hep-th/0212095.
}

\lref\BenaTN{ I.~Bena, S.~de Haro and R.~Roiban, ``Generalized
Yukawa couplings and matrix models,'' arXiv:hep-th/0212083.
}

\lref\OhtaRD{ K.~Ohta, ``Exact mesonic vacua from matrix models,''
arXiv:hep-th/0212025.
}

\lref\OokouchiBE{ Y.~Ookouchi, ``N = 1 gauge theory with flavor
from fluxes,'' arXiv:hep-th/0211287.
}

\lref\FengZB{ B.~Feng, ``Seiberg duality in matrix model,''
arXiv:hep-th/0211202.
}

\lref\BenaUA{ I.~Bena, R.~Roiban and R.~Tatar, ``Baryons,
boundaries and matrix models,'' arXiv:hep-th/0211271.
}

\lref\NaculichHR{ S.~G.~Naculich, H.~J.~Schnitzer and N.~Wyllard,
``Matrix model approach to the N = 2 U(N) gauge theory with matter
in the  fundamental representation,'' arXiv:hep-th/0211254.
}

\lref\ArgurioHK{ R.~Argurio, V.~L.~Campos, G.~Ferretti and
R.~Heise, ``Baryonic corrections to superpotentials from
perturbation theory,'' arXiv:hep-th/0211249.
}

\lref\IntriligatorAU{ K.~A.~Intriligator and N.~Seiberg,
``Lectures on supersymmetric gauge theories and electric-magnetic
duality,'' Nucl.\ Phys.\ Proc.\ Suppl.\  {\bf 45BC}, 1 (1996)
[arXiv:hep-th/9509066].
}

\lref\FengYF{ B.~Feng and Y.~H.~He, ``Seiberg duality in matrix
models. II,'' arXiv:hep-th/0211234.
}

\lref\KapustinNB{ A.~Kapustin, ``The Coulomb branch of N = 1
supersymmetric gauge theory with adjoint  and fundamental
matter,'' Phys.\ Lett.\ B {\bf 398}, 104 (1997)
[arXiv:hep-th/9611049].
}

\lref\SeibergAJ{ N.~Seiberg and E.~Witten, ``Monopoles, duality
and chiral symmetry breaking in N=2 supersymmetric QCD,'' Nucl.\
Phys.\ B {\bf 431}, 484 (1994) [arXiv:hep-th/9408099].
}

\lref\TachikawaWK{ Y.~Tachikawa, ``Derivation of the Konishi
anomaly relation from Dijkgraaf-Vafa with  (bi-)fundamental
matters,'' arXiv:hep-th/0211189.
}

\lref\CachazoZK{ F.~Cachazo, N.~Seiberg and E.~Witten, ``Phases of
N = 1 supersymmetric gauge theories and matrices,''
arXiv:hep-th/0301006.
}

\lref\DemasureSC{ Y.~Demasure and R.~A.~Janik, ``Effective matter
superpotentials from Wishart random matrices,''
arXiv:hep-th/0211082.
}

\lref\BenaKW{ I.~Bena and R.~Roiban, ``Exact superpotentials in N
= 1 theories with flavor and their matrix  model formulation,''
arXiv:hep-th/0211075.
}

\lref\SuzukiGP{ H.~Suzuki, ``Perturbative derivation of exact
superpotential for meson fields from  matrix theories with one
flavour,'' arXiv:hep-th/0211052.
}

\lref\McGreevyYG{ J.~McGreevy, ``Adding flavor to
Dijkgraaf-Vafa,'' arXiv:hep-th/0211009.
}

\lref\ArgurioXV{ R.~Argurio, V.~L.~Campos, G.~Ferretti and
R.~Heise, ``Exact superpotentials for theories with flavors via a
matrix integral,'' arXiv:hep-th/0210291.
}

\lref\BerensteinSN{ D.~Berenstein, ``Quantum moduli spaces from
matrix models,'' arXiv:hep-th/0210183.
}

\lref\HollowoodZK{ T.~J.~Hollowood and T.~Kingaby, ``The phase
structure of mass-deformed SU(2) x SU(2) quiver theory,''
arXiv:hep-th/0210096.
}

\lref\agmoo{ O.~Aharony, S.~S.~Gubser, J.~M.~Maldacena, H.~Ooguri
and Y.~Oz, ``Large N field theories, string theory and gravity,''
Phys.\ Rept.\  {\bf 323}, 183 (2000) arXiv:hep-th/9905111.}

\lref\NovikovEE{V.~A.~Novikov, M.~A.~Shifman, A.~I.~Vainshtein and
V.~I.~Zakharov, ``Instanton Effects In Supersymmetric Theories,''
Nucl.\ Phys.\ B {\bf 229}, 407 (1983).
}

\lref\ArgyresJJ{ P.~C.~Argyres and M.~R.~Douglas, ``New phenomena
in SU(3) supersymmetric gauge theory,'' Nucl.\ Phys.\ B {\bf 448},
93 (1995) [arXiv:hep-th/9505062].
}

\lref\ArgyresXN{ P.~C.~Argyres, M.~Ronen Plesser, N.~Seiberg and
E.~Witten, ``New N=2 Superconformal Field Theories in Four
Dimensions,'' Nucl.\ Phys.\ B {\bf 461}, 71 (1996)
[arXiv:hep-th/9511154].
}

\lref\DijkgraafXD{ R.~Dijkgraaf, M.~T.~Grisaru, C.~S.~Lam, C.~Vafa
and D.~Zanon, ``Perturbative computation of glueball
superpotentials,'' arXiv:hep-th/0211017.
}

\lref\SeibergRS{ N.~Seiberg and E.~Witten, ``Electric - magnetic
duality, monopole condensation, and confinement in N=2
supersymmetric Yang-Mills theory,'' Nucl.\ Phys.\ B {\bf 426}, 19
(1994) [Erratum-ibid.\ B {\bf 430}, 485 (1994)]
[arXiv:hep-th/9407087].
}

\lref\kovner{A. Kovner and M. Shifman, ``Chirally Symmetric Phase
Of Supersymmetric Gluodynamics,'' Phys. Rev. {\bf D56} (1997)
2396, hep-th/9702174.}

\lref\konishione{ K.~Konishi, ``Anomalous Supersymmetry
Transformation Of Some Composite Operators In Sqcd,'' Phys.\
Lett.\ B {\bf 135}, 439 (1984).
}

\lref\konishitwo{ K.~i.~Konishi and K.~i.~Shizuya, ``Functional
Integral Approach To Chiral Anomalies In Supersymmetric Gauge
Theories,'' Nuovo Cim.\ A {\bf 90}, 111 (1985).
}

\lref\FradkinDV{ E.~H.~Fradkin and S.~H.~Shenker, ``Phase Diagrams
Of Lattice Gauge Theories With Higgs Fields,'' Phys.\ Rev.\ D {\bf
19}, 3682 (1979).
}

\lref\BanksFI{ T.~Banks and E.~Rabinovici, ``Finite Temperature
Behavior Of The Lattice Abelian Higgs Model,'' Nucl.\ Phys.\ B
{\bf 160}, 349 (1979).
}

\lref\arkmur{N. Arkani-Hamed and H. Murayama, hep-th/9707133.}

\lref\bipz{ E.~Brezin, C.~Itzykson, G.~Parisi and J.~B.~Zuber,
``Planar Diagrams,'' Commun.\ Math.\ Phys.\  {\bf 59}, 35 (1978).
}

\lref\BershadskyCX{ M.~Bershadsky, S.~Cecotti, H.~Ooguri and
C.~Vafa, ``Kodaira-Spencer theory of gravity and exact results for
quantum string amplitudes,'' Commun.\ Math.\ Phys.\  {\bf 165},
311 (1994) [arXiv:hep-th/9309140].
}

\lref\mmreview{ P.~Ginsparg and G.~W.~Moore, ``Lectures On 2-D
Gravity And 2-D String Theory,'' arXiv:hep-th/9304011.
}

\lref\gorsky{ A.~Gorsky, ``Konishi anomaly and N = 1 effective
superpotentials from matrix models,'' arXiv:hep-th/0210281.
}

\lref\CachazoJY{ F.~Cachazo, K.~A.~Intriligator and C.~Vafa, ``A
large N duality via a geometric transition,'' Nucl.\ Phys.\ B {\bf
603}, 3 (2001) [arXiv:hep-th/0103067].
}

\lref\KutasovVE{ D.~Kutasov, ``A Comment on duality in N=1
supersymmetric nonAbelian gauge theories,'' Phys.\ Lett.\ B {\bf
351}, 230 (1995) [arXiv:hep-th/9503086].
}

\lref\FerrariJP{ F.~Ferrari, ``On exact superpotentials in
confining vacua,'' arXiv:hep-th/0210135.
}

\lref\WittenXI{ E.~Witten, ``The Verlinde Algebra And The
Cohomology Of The Grassmannian,'' arXiv:hep-th/9312104, and in
{\it Quantum Fields And Strings: A Course For Mathematicians}, ed.
P. Deligne et. al. (American Mathematical Society, 1999), vol. 2,
pp. 1338-9.
}

\lref\SeibergPQ{ N.~Seiberg, ``Electric - magnetic duality in
supersymmetric nonAbelian gauge theories,'' Nucl.\ Phys.\ B {\bf
435}, 129 (1995) [arXiv:hep-th/9411149].
}

\lref\IntriligatorID{ K.~A.~Intriligator and N.~Seiberg,
``Duality, monopoles, dyons, confinement and oblique confinement
in supersymmetric SO(N(c)) gauge theories,'' Nucl.\ Phys.\ B {\bf
444}, 125 (1995) [arXiv:hep-th/9503179].
}

\lref\FujiWD{ H.~Fuji and Y.~Ookouchi, ``Comments on effective
superpotentials via matrix models,'' arXiv:hep-th/0210148.
}

\lref\KutasovVE{ D.~Kutasov, ``A Comment on duality in N=1
supersymmetric nonAbelian gauge theories,'' Phys.\ Lett.\ B {\bf
351}, 230 (1995) [arXiv:hep-th/9503086].
}

\lref\KutasovNP{ D.~Kutasov and A.~Schwimmer, ``On duality in
supersymmetric Yang-Mills theory,'' Phys.\ Lett.\ B {\bf 354}, 315
(1995) [arXiv:hep-th/9505004].
}

\lref\KutasovSS{ D.~Kutasov, A.~Schwimmer and N.~Seiberg, ``Chiral
Rings, Singularity Theory and Electric-Magnetic Duality,'' Nucl.\
Phys.\ B {\bf 459}, 455 (1996) [arXiv:hep-th/9510222].
}

\lref\DijkgraafFC{ R.~Dijkgraaf and C.~Vafa, ``Matrix models,
topological strings, and supersymmetric gauge theories,''
arXiv:hep-th/0206255.
}

\lref\DijkgraafVW{ R.~Dijkgraaf and C.~Vafa, ``On geometry and
matrix models,'' arXiv:hep-th/0207106.
}

\lref\OoguriGX{ H.~Ooguri and C.~Vafa, ``Worldsheet derivation of
a large N duality,'' Nucl.\ Phys.\ B {\bf 641}, 3 (2002)
[arXiv:hep-th/0205297].
}

\lref\Macdonald{I.G.~Macdonald, ``The volume of a compact Lie
group,'' Invent. Math. {\bf 56}, 93 (1980). }

\lref\DijkgraafDH{ R.~Dijkgraaf and C.~Vafa, ``A perturbative
window into non-perturbative physics,'' arXiv:hep-th/0208048.
}

\lref\IntriligatorJR{ K.~A.~Intriligator, R.~G.~Leigh and
N.~Seiberg, ``Exact superpotentials in four-dimensions,'' Phys.\
Rev.\ D {\bf 50}, 1092 (1994) [arXiv:hep-th/9403198].
}

\lref\SeibergBZ{ N.~Seiberg, ``Exact results on the space of vacua
of four-dimensional SUSY gauge theories,'' Phys.\ Rev.\ D {\bf
49}, 6857 (1994) [arXiv:hep-th/9402044].
}

\lref\CachazoRY{ F.~Cachazo, M.~R.~Douglas, N.~Seiberg and
E.~Witten, ``Chiral rings and anomalies in supersymmetric gauge
theory,'' arXiv:hep-th/0211170.
}

\lref\tyu{A.~Hanany and Y.~Oz, ``On the quantum moduli space of
vacua of N=2 supersymmetric SU(N(c)) gauge theories,'' Nucl.\
Phys.\ B {\bf 452}, 283 (1995) [arXiv:hep-th/9505075].
P.~C.~Argyres, M.~R.~Plesser and A.~D.~Shapere, ``The Coulomb
phase of N=2 supersymmetric QCD,'' Phys.\ Rev.\ Lett.\  {\bf 75},
1699 (1995) [arXiv:hep-th/9505100].
J.~A.~Minahan and D.~Nemeschansky, ``Hyperelliptic curves for
supersymmetric Yang-Mills,'' Nucl.\ Phys.\ B {\bf 464}, 3 (1996)
[arXiv:hep-th/9507032].
I.~M.~Krichever and D.~H.~Phong, ``On the integrable geometry of
soliton equations and N = 2  supersymmetric gauge theories,'' J.\
Diff.\ Geom.\  {\bf 45}, 349 (1997) [arXiv:hep-th/9604199].
E.~D'Hoker, I.~M.~Krichever and D.~H.~Phong, ``The effective
prepotential of N = 2 supersymmetric SU(N(c)) gauge  theories,''
Nucl.\ Phys.\ B {\bf 489}, 179 (1997) [arXiv:hep-th/9609041].
}

\lref\DijkgraafPP{ R.~Dijkgraaf, S.~Gukov, V.~A.~Kazakov and
C.~Vafa, ``Perturbative analysis of gauged matrix models,''
arXiv:hep-th/0210238.
}

\lref\GopakumarWX{ R.~Gopakumar, ``${\cal N}=1$ Theories and a
Geometric Master Field,'' arXiv:hep-th/0211100.
}

\lref\Schnitzer{S.G.~ Naculich, H.J.~ Schnitzer and N.~Wyllard,
``The $\CN=2$ $U(N)$ gauge theory prepotential and periods from a
perturbative matrix model calculation,'' arXiv:hep-th/0211123.}

\lref\SeibergVC{ N.~Seiberg, ``Naturalness versus supersymmetric
nonrenormalization theorems,'' Phys.\ Lett.\ B {\bf 318}, 469
(1993) [arXiv:hep-ph/9309335].
}

\lref\DouglasNW{ M.~R.~Douglas and S.~H.~Shenker, ``Dynamics of
SU(N) supersymmetric gauge theory,'' Nucl.\ Phys.\ B {\bf 447},
271 (1995) [arXiv:hep-th/9503163].
}

\lref\CachazoPR{ F.~Cachazo and C.~Vafa, ``N = 1 and N = 2
geometry from fluxes,'' arXiv:hep-th/0206017.
}

\lref\IntriligatorJR{ K.~A.~Intriligator, R.~G.~Leigh and
N.~Seiberg, ``Exact superpotentials in four-dimensions,'' Phys.\
Rev.\ D {\bf 50}, 1092 (1994) [arXiv:hep-th/9403198].
}

\lref\DijkgraafXD{ R.~Dijkgraaf, M.~T.~Grisaru, C.~S.~Lam, C.~Vafa
and D.~Zanon, ``Perturbative Computation of Glueball
Superpotentials,'' arXiv:hep-th/0211017.
}
\lref\superspace{ S.~J.~Gates, M.~T.~Grisaru, M.~Rocek and
W.~Siegel, ``Superspace, Or One Thousand And One Lessons In
Supersymmetry,'' Front.\ Phys.\  {\bf 58}, 1 (1983)
[arXiv:hep-th/0108200].
}

\lref\nicolai{ H.~Nicolai, ``On A New Characterization Of Scalar
Supersymmetric Theories,'' Phys.\ Lett.\ B {\bf 89}, 341 (1980).
}

\lref\migdal{ A.~A.~Migdal, ``Loop Equations And 1/N Expansion,''
Phys.\ Rept.\  {\bf 102}, 199 (1983).
}

\lref\FengIS{ B.~Feng, ``Note on matrix model with massless
flavors,'' arXiv:hep-th/0212274.
}

\lref\RoibanUQ{ R.~Roiban, R.~Tatar and J.~Walcher, ``Massless
flavor in geometry and matrix models,'' arXiv:hep-th/0301217.
}

\lref\BalasubramanianTV{ V.~Balasubramanian, B.~Feng, M.~x.~Huang
and A.~Naqvi, ``Phases of N=1 Supersymmetric Gauge Theories with
Flavors,'' arXiv:hep-th/0303065.
}

\lref\FengEG{ B.~Feng, ``Note on Seiberg Duality in Matrix
Model,'' arXiv:hep-th/0303144.
}

\lref\BenaVK{ I.~Bena, H.~Murayama, R.~Roiban and R.~Tatar,
``Matrix Model Description of Baryonic Deformations,''
arXiv:hep-th/0303115.
}

\lref\staudacher{ M.~Staudacher, ``Combinatorial solution of the
two matrix model,'' Phys.\ Lett.\ B {\bf 305}, 332 (1993)
[arXiv:hep-th/9301038].
}

\lref\voiculescu{{\it Free Probability Theory}, ed. D. Voiculescu,
pp. 21--40, AMS, 1997.}

\lref\CeresoleZS{ A.~Ceresole, G.~Dall'Agata, R.~D'Auria and
S.~Ferrara, ``Spectrum of type IIB supergravity on AdS(5) x T(11):
Predictions on N  = 1 SCFT's,'' Phys.\ Rev.\ D {\bf 61}, 066001
(2000) [arXiv:hep-th/9905226].
}

\def\tilde{\widetilde}
\def\CN{{\cal N}}
\def\CM{{\cal M}}
\def\h{\zeta}
\def\t{\tau}
\def\R{{ \Sigma}}
\def\b{B}
\def\B{{\hat B}}

\def\rb{{\hat B}^{r}}
\def\H{{\cal H}}
\def\G{{\cal G}}
\def\tp{{1\over 2\pi i}}
\def\q{q}
\def\p{p}

\lref\fark{H.M. Farkas and I. Kra. {\it Riemann Surfaces}, 2nd
ed., Springer-Verlag, 1992.}

\newbox\tmpbox\setbox\tmpbox\hbox{\abstractfont
}
 \Title{\vbox{\baselineskip12pt\hbox to\wd\tmpbox{\hss
 hep-th/0303207} }}
 {\vbox{\centerline{Chiral Rings and Phases}
 \smallskip
 \centerline{of Supersymmetric Gauge Theories}
 }}
\smallskip
\centerline{Freddy Cachazo, Nathan Seiberg and Edward Witten}
\smallskip
\bigskip
\centerline{School of Natural Sciences, Institute for Advanced
Study, Princeton NJ 08540 USA}
\bigskip
\vskip 1cm
 \noindent
We solve for the expectation values of chiral operators in
supersymmetric $U(N)$ gauge theories with matter in the adjoint,
fundamental and anti-fundamental representations.  A simple
geometric picture emerges involving a description by a meromorphic
one-form on a Riemann surface.  The equations of motion are
equivalent to a condition on the integrality of periods of this
form.  The solution indicates that all semiclassical phases with
the same number of $U(1)$ factors are continuously connected.

\Date{March 2003}
%

%
%

\newsec{Introduction}

In the previous century, it became clear that supersymmetric field
theories exhibit rich dynamics which is amenable to exact analysis
(for a review, see e.g.\ \IntriligatorAU).  A renaissance of this
subject has recently been stimulated by the work of Dijkgraaf and
Vafa \DijkgraafDH, who, motivated by earlier developments
\refs{\BershadskyCX\GopakumarKI\VafaWI\CachazoJY\CachazoPR
\DijkgraafFC-\DijkgraafVW},  conjectured an interesting relation
between SUSY gauge theories and matrix models.  Many authors have
added to this framework matter in the fundamental representation
\refs{\HollowoodZK\BerensteinSN
\ArgurioXV\McGreevyYG\SuzukiGP\BenaKW\DemasureSC
\TachikawaWK\ArgurioHK\NaculichHR\BenaUA
\FengZB\FengYF\OokouchiBE\OhtaRD\BenaTN\HofmanBI\SeibergJQ
\FengIS\RoibanUQ\BalasubramanianTV\BenaVK-\FengEG}.  Here we will
continue the investigation in our previous papers
\refs{\CachazoRY,\SeibergJQ,\CachazoZK}, in which these theories
have been studied by focusing on the chiral ring and the relations
which follow from the anomaly. As we will see, these relations can
be explicitly solved.

We will study a supersymmetric $U(N)$ gauge theory with chiral
superfields consisting of a multiplet $\Phi$ in the adjoint
representation, plus $N_f$ copies of the fundamental
representation  (quarks, denoted $Q^f$, $f=1,\dots,N_f$) and  the
anti-fundamental representation (anti-quarks, denoted $\tilde
Q_f$). We take the superpotential to be
 \eqn\treesupai{W_{tree}=\Tr W(\Phi) + \tilde Q_{\tilde f}
 m^{\tilde f}_f (\Phi) Q^f,}
where we suppressed the color indices in the second term, and
$W(\Phi)$ and $m^{\tilde f}_f (\Phi) $ are polynomials.
Classically the theory has several vacua.  First, the fundamental
and anti-fundamental fields $Q$ and $\tilde Q$ can vanish while
$\Phi$ is a diagonal matrix whose eigenvalues are at the
stationary points of $W(\Phi)$.  In these vacua, the quarks and
anti-quarks are massive, and the microscopic $U(N)$ gauge symmetry
is broken to $\prod_i U(N_i)$ with $\sum_i N_i=N$. The second kind
of classical vacua involve nonzero expectation values of $Q$ and
$\tilde Q$. Here the gauge symmetry is broken to $\prod_i U(N_i)$
with $\sum_i N_i< N$.  In the quantum theory the unbroken $\prod_i
SU(N_i)$ confines and the low energy spectrum includes a number of
$U(1)$ multiplets.

Our main tool in analyzing the dynamics is the chiral ring of the
theory.  Interesting bosonic operators in this ring are the gauge
invariant observables \refs{\CachazoRY,\SeibergJQ}
 \eqn\chiralopfi{\eqalign{
 T(z)&= \Tr {1\over z-\Phi} \cr
 R(z)&= -{1 \over 32 \pi^2} \Tr {W_\alpha W^\alpha \over
 z-\Phi}\cr
 M(z)_{\tilde f}^f &= \tilde Q_{\tilde f}  {1\over
 z-\Phi}Q^f.}}
Classically, $z$ takes values in the complex plane, and these
observables have simple poles where $z$ equals an eigenvalue of $\Phi$.
  In the quantum theory, some of the
poles become cuts and although an {\it a priori} reason is not really
understood, it turns out that $z$ can be analytically continued through
the cuts to a second sheet.  The two-sheeted plane is a Riemann
surface $\Sigma$ which is described by the equation
 \eqn\riee{y^2=W'(z)^2 + f(z)}
where $R(z)$ of \chiralopfi\ is given by $(W'(z)-y(z))/2$.  In
\riee, $W'(z)$ and $f(z)$ are polynomials of degree $n$ and $n-1$
respectively, and \riee\ describes a genus $n-1$ Riemann surface.

The exact solution corresponds to solving for the polynomial
$f(z)$, which depends only on the vacuum of the theory. However,
it is often useful to consider the ``off-shell'' theory for
arbitrary values of the $n$ coefficients of this polynomial.
Alternatively, we can parametrize the solution by the $n$ periods
 \eqn\Siin{S_i ={1\over 2\pi i} \oint_{A_i} R(z) dz,}
where $A_i$ are $A$-cycles of $\Sigma$.  Semiclassically, these
$A$-cycles originate from the poles of the classical theory,
which quantum mechanically became cuts in the $z$
plane. Important objects in our discussion are the integers
 \eqn\Niin{N_i ={1\over 2\pi i} \oint_{A_i} T(z)dz,}
which are the ranks of the various unbroken gauge groups.

In the following sections we describe the theory in more detail
and solve off-shell,   using
relations in the chiral ring, for $T(z)$ and $M(z)$, which were defined in
 \chiralopfi.  One of the surprises  we
encounter is that $T(z)$ has a very simple analytic structure: its
only singularities are simple poles with integer residues. This is
particularly surprising for the poles in the second sheet which
are not even visible classically. We then view $T(z)dz$ as a
one-form on $\Sigma$ and explore its periods.  Integrality of the
periods in \Niin\ plus ``modular invariance'' (invariance under
exchanges of the periods that arise when parameters in the
superpotential are varied) suggests that all periods of $T(z)dz$
might be integers. In particular, the other compact periods of $T$
 \eqn\biin{b_i =-{1\over 2\pi i} \oint_{\b_i} T(z) dz,}
 which were introduced and interpreted physically in \CachazoZK, are integers
on-shell.  The constraints on the analytic structure of
$T(z)dz$ and its periods allow us to find an explicit expression
for it.

We then compute the off-shell effective superpotential as a
function of $S_i$ using the matrix model.  It is satisfying to
find out that the equations of motion of $S_i$ which are derived
from this superpotential coincide with the previously imposed
geometric condition about the periods of $T(z)dz$.  This result is
consistent with the fact that \biin\ is satisfied only on-shell
and that imposing \biin\ allows us to completely solve for our
observables.

In \CachazoZK, we explored the phases of the theory with $\Phi$
only (and no quarks) as the parameters in the tree level
superpotential are varied.  Some of the phases of the theory were
distinguished by the behavior of Wilson and 't Hooft loops which
probe the different kinds of confinement.  Here, since the theory
includes fundamental matter fields, there is no real confinement,
and we might expect that there are no phase transitions
\refs{\FradkinDV,\BanksFI}.  We will use the term
``pseudo-confining'' to describe a phase in which $Q$ and $\tilde
Q$ have no classical expectation values and the individual low
energy $SU(N_i)$ theories are confining, though the microscopic
$U(N)$ theory is not. The result of \refs{\FradkinDV,\BanksFI} is
the possibility of smooth interpolation between Higgs and
pseudo-confinement.

This subject has also been  discussed in $\CN=2$ supersymmetric
theories with matter, first with $SU(2)$ gauge theories
\SeibergAJ, and later with other gauge groups \tyu. It was shown
that the massless particles at special points in the moduli space
can be interpreted either as elementary quarks or as magnetic
monopoles. More explicitly, by varying the bare mass of the
quarks, one can continuously interpolate between a limit in which
it is more natural to interpret the massless particles as
electrically charged to a limit in which it is more natural to
interpret them as magnetic. When $\CN=2$ is broken to $\CN=1$ by a
mass term for the adjoint fields, these quarks/monopoles condense.
In one limit, where the condensed particles are electrically
charged, it is natural to interpret the condensation as a Higgs
mechanism. In the other limit, where the condensed objects are
interpreted as magnetic monopoles, the condensation looks in the
low energy pure $SU(2)$ theory as confinement, but in the full
theory including the massive quarks, there is no precise order
parameter for confinement, and we will call this limit
pseudo-confining. Therefore, by changing the bare mass of the
quarks, one can continuously interpolate between Higgs and
pseudo-confinement.

The situation we have just described is a special case of a more
general story we will describe below.  We will show that as we
vary the parameters in $m(\Phi)$, we can connect almost all the
classical limits of the theory with the same number of $U(1)$
factors in the low energy spectrum.

The poles of $T(z)$ play a crucial role in the interpolation from
one classical limit to another.  A pole in the first sheet
represents a classical nonzero expectation value of $Q$, i.e., a
Higgs phenomenon.  A pole in the second sheet corresponds to an
expectation value of $Q$ that arises only from quantum
corrections.  In the quantum theory, the Riemann surface is smooth
and there is no physical boundary between the first and the second
sheet. The poles can continuously move from one sheet to another,
a phenomenon which represents the continuous interpolation between
the Higgs mechanism and pseudo-confinement.

By analogy with \CachazoZK, we explore the various classical
limits of each branch by performing modular transformations on
$\Sigma$.  Unlike the situation in \CachazoZK, because of the
presence of the fundamental matter, $T$ has poles and $\Sigma$
should be thought of as a Riemann surface with punctures. This
makes the relevant modular group bigger.  Using appropriate
modular transformations, we can connect all values of $b_i$ with
fixed $N_i$; we can also connect most values of $N_i$. Therefore,
the quantum theory has many fewer phases than the classical
theory.

\newsec{Preliminaries}

We consider an ${\cal N}=1$ supersymmetric $U(N)$ gauge theory
with chiral superfields consisting of an adjoint multiplet
    $\Phi$, $N_f$ fundamentals $Q^f$, and
$N_f$ anti-fundamentals $\tilde Q_{\tilde f}$ ($f$ and $\tilde f$
are the flavor indices). As in the introduction,
the tree level superpotential is
 \eqn\treesupa{W_{tree}=\Tr W(\Phi) + \tilde Q_{\tilde f}
 m^{\tilde f}_f (\Phi) Q^f.}
  The function $W$ and the
matrix $m$ are taken to be polynomials
 \eqn\coeffmg{\eqalign{
 W(z)&=\sum_{k=0}^n {1\over k+1} g_k z^{k+1} \cr
 m_f^{\tilde f}(z) &=\sum_{k=1}^{l+1}  m_{f, k}^{\tilde
 f}z^{k-1}.}}

It is convenient to define the polynomial
 \eqn\Bdef{B(z)=\det m(z).}
We denote its degree by $L$ and its roots by $z_I$ ($I=1,...,L$).
We will see from the solution of the theory that some of its
observables depend only on $B(z)$ and not on the details of the
matrix $m_f^{\tilde f}(z)$. $L$ will play the role of an effective
number of flavors. We assume that $m_f^{\tilde f}(z)$ is sufficiently generic
so that
 $B(z)$ does not have double roots, and none of its
roots coincide with those of $W'(z)$.

Some special cases are the following:
 \item{1.} No flavors.  This is the basic example which is most widely
studied.  When $W(\Phi)=0$, this theory has $\CN=2$
 supersymmetry.
 \item{2.} No $\Phi$ or equivalently $W(\Phi)=M\Phi^2$ with large $M$.
  This theory is $\CN=1$ SQCD.
 \item{3.} $m_f^{\tilde f}(z) = (\sqrt{2}\; z+m_f)\delta _f^{\tilde f}$.
 Here the theory without $W(\Phi)$ has $\CN=2$ supersymmetry.
 \item{4.} $m_f^{\tilde f}(z) = z^l\delta _f^{\tilde f}$ with
 $W(\Phi)=0$.  This theory was studied in \KapustinNB.

 Let us examine the vacua of the theory in the semiclassical
approximation.  We parametrize the superpotential of the adjoint
field as
 \eqn\superpotai{W'(z)= g_n\prod_{i=1}^n (z-a_i)}
in terms of its stationary points $a_i$.  A first set of vacua,
which we will refer to loosely as ``pseudo-confining vacua,''
arise from
 \eqn\Phiex{\eqalign{\langle Q\rangle =& \langle \tilde Q \rangle
 =0\cr
 \langle \Phi \rangle =& \pmatrix{
 a_1&&&&&&&&&\cr
 &.  &&&&&&&&\cr
 &&.  &&&&&&&\cr
 &&&a_2&&&&&&\cr
 &&&&.  &&&&&\cr
 &&&&&.  &&&&\cr
 &&&&&&a_3&&&\cr
 &&&&&&&.  &&\cr
 &&&&&&&&.  &\cr
 &&&&&&&&&a_n\cr}}}
where $a_i$ occurs $N_i$ times ($\sum_i N_i = N$).  Here, the
fundamental $U(N)$ gauge group is broken to $\prod_i U(N_i)$.  At
low energies the $ SU(N_i)$ confine, leading to $\prod_i N_i$
vacua in which the low energy gauge group is $U(1)^k$.  $k$ can be
less than $n$ if some of the $a_i$ do not appear in \Phiex\ and
the corresponding $N_i$ are equal to zero.

There are also Higgs vacua in which $Q$ and $\tilde Q$ have a
vacuum expectation value at the classical level, and one of the
diagonal elements of $\langle \Phi\rangle$ is equal to a  zero of
$B(z)$, say $z_1$.
 In explaining how this works,
 in order not to clutter the equations, we consider the simple
case of $N_f=1$. The choice
 \eqn\Phiexh{\eqalign{
 \langle Q\rangle = &\pmatrix{h&0&&.&&.&&.&&0}\cr
 \langle \tilde Q\rangle = &\pmatrix{\tilde h&0&&.&&.&&.&&0}\cr
 \langle \Phi \rangle =&\pmatrix{
 z_1&&&&&&&&&\cr
 &a_1  &&&&&&&&\cr
 &&.  &&&&&&&\cr
 &&&a_2&&&&&&\cr
 &&&&.  &&&&&\cr
 &&&&&.  &&&&\cr
 &&&&&&a_3&&&\cr
 &&&&&&&.  &&\cr
 &&&&&&&&.  &\cr
 &&&&&&&&&a_n\cr}}}
gives a classical vacuum when
 \eqn\qdet{\tilde h h = - {W'(z_1)\over B'(z_1)}.}
In such a vacuum, the fundamental $U(N)$ gauge group is broken to
$\prod_i U(N_i)$ with $\sum_i N_i = N-1$.

We cannot have two different eigenvalues of $\langle \Phi
\rangle$, say $\langle \Phi^1{}_1\rangle$ and $\langle \Phi
^2{}_2\rangle$, equal to $z_1$. In this case $\langle
Q_{1,2}\rangle$ and $\langle \tilde Q^{1,2}\rangle$ must be
nonzero, but then the equation of motion of $\Phi^2{}_1$ is not
satisfied.

More generally, for each zero $z_I$ of $B(z)$, we let $r_I$ denote
the number  of eigenvalues of  $ \Phi $ that are equal to $z_I$.
In a classical vacuum, each $r_I$ can be either zero or one.
Clearly,
 \eqn\sumrI{0\le \sum_{I=1}^L r_I  \le L,}
and
 \eqn\NLine{
 N =\sum_{I=1}^L r_I +\sum_{i=1}^n N_i.}
We will refer to vacua with nonzero $r_I$ as ``Higgs vacua.''
Again, at low energies $\prod_i SU(N_i)$ confine and lead to
$\prod_i N_i$ vacua with gauge group $U(1)^k$.  We will see below
that in the quantum theory the pseudo-confining vacua \Phiex\ are
continuously connected to the Higgs vacua \Phiexh.

We will be interested in the chiral operators
\refs{\CachazoRY,\SeibergJQ}
 \eqn\chiralopf{\eqalign{
 T(z)&= \Tr {1\over z-\Phi} \cr
 w_\alpha(z)&= {1 \over 4\pi} \Tr {W_\alpha\over z-\Phi} \cr
 R(z)&= -{1 \over 32 \pi^2} \Tr {W_\alpha W^\alpha \over
 z-\Phi}\cr
 M(z)_{\tilde f}^f &= \tilde Q_{\tilde f}  {1\over
 z-\Phi}Q^f\cr}}
These composite operators are defined initially in terms of a
power series in ${1\over z}$ around infinity, but turn out to have
an interesting analytic continuation. All these operators are
$\CO(1/z)$ for $z\to\infty$.

First, let us examine these observables in the classical theory.
For $T$, we easily find
 \eqn\Tclz{T_{cl}(z)=\sum_i {N_i \over z-a_i} + \sum_I {r_I
 \over z-z_I}.}
 We will view $T(z)dz$ as a one-form.  Classically, its only singularities
 are
simple poles at $a_i$, $z_I$ and infinity with residues $N_i$,
$r_I$ and $-N$, respectively.  Generalizing \qdet, we find
 \eqn\resMc{ M_{cl}(z) = - \sum_{I=1}^L {r_IW'(z_I)  \over z- z_I }
 \tp\oint _{z_I} {1 \over m(x)} dx,}
i.e.\ there are poles only at $z_I$ with $r_I=1$.  In contrast to
poles of $T$, the poles of $M_{cl}$ have residues that depend on
the details of the interactions. The classical limit of $R(z)$ is
zero, since it is proportional to a fermion bilinear.

Now let us give a preview of what will occur in the quantum
theory.  Quantum mechanically, when the fermion bilinear in the
numerator of $R(z)$ gets an expectation value, one might expect
$R(z)$ to develop a singularity at $a_i$ and/or (in a Higgs
vacuum) $z_I$ because the denominator of $R(z)$ is singular when
$z$ approaches an eigenvalue of $\Phi$.  We will see that actually
$R(z)$ has no singularities at $z_I$, while the singularities of
$R(z)$  associated with $a_i$ are replaced by cuts that shrink to
$a_i$ in a weak coupling limit; we let $A_i$ denote a cycle that
circles counterclockwise once around the $i^{th}$ such cut.
Because of the cuts, $R(z)$ is naturally defined on a Riemann
surface $\Sigma$ that is a double cover of the $z$-plane.  Any
value of $z$, such as $z=z_I$, really corresponds to a pair of
points on $\Sigma$; we denote the points with $z=z_I$ as $\q_I$,
which is on the first sheet and visible semiclassically, and
$\tilde \q_I$, which is on the second sheet and invisible
semiclassically.

$M$ and $T$ are also naturally defined on this double cover, and
as functions on $\Sigma$, their only singularities turn out to be
simple poles. The classical pole of $T$ at $a_i$ with residue
$N_i$ is replaced by a quantum statement $\tp\oint_{A_i} T(z) dz =
N_i$; in effect, these poles disappear when the points $a_i$ are
replaced by cuts and the $z$-plane by its double cover $\Sigma$.
The other classical singularities of $T$ are the poles at those
$z_I$ with $r_I=1$; these are also the locations of the only
classical singularities of $M$. What happens to these
singularities quantum mechanically? They remain as simple poles
and, roughly speaking, they do not move from their classical
locations at $z_I$.  To be more precise, these poles appear
quantum mechanically at the points $\q_I$ that lie above $z_I$ on
the first sheet of $\Sigma$.  The residues of $T(z) dz$ at these
poles are equal to 1 quantum mechanically just as classically;
this actually follows by an argument similar to one in \CachazoRY.
Indeed, letting $C_I$ denote a small contour around $\q_I$, the
contour integral \eqn\polyo{{1\over 2\pi
i}\oint_{C_I}T(z)dz={1\over 2\pi i}\oint_{C_I}\Tr{1\over
z-\Phi}dz,} is really a $c$-number, not subject to quantum
fluctuations, since it is equal to the number of eigenvalues of
$\Phi$ enclosed by the contour.  If the only singularity of $T(z)$
enclosed by $C_I$ is a simple pole, the residue must hence be one.

On the other hand, the residue of the pole of $M$ does receive
quantum corrections; it is not protected by any such argument,
because of the quark operators in the numerator of $M$. In
addition to the singularities at $\q_I$, $T$ and $M$ will turn out
in general to have poles, crucial in the consistency of the whole
picture, at the points $\tilde \q_I$ on the second sheet. The
meaning of these poles is somewhat mysterious from a
semi-classical point of view, as the second sheet is invisible
classically.  Quantum mechanically, they reflect the fact that the
Higgs phases can be reached from the pseudo-confining phase by
analytic continuation.

\bigskip\noindent{\it Anomaly Equations}

To verify and extend these statements, we will use the relations
obeyed by the operators \chiralopf\ in the chiral ring. These
operators satisfy the anomaly equations
\refs{\CachazoZK,\SeibergJQ}
  \eqn\anoeff{\eqalign{
 &\left[W'(z)T(z)\right]_- + \tr \left[m'(z)M(z) \right]_- =
 2R(z)T(z) +w_\alpha(z) w^\alpha(z)\cr
 &\left[W'(z) w_\alpha(z)\right]_-  = 2R(z)w_\alpha(z)\cr
 &\left[W'(z) R(z)\right]_-  = R(z)^2\cr
 & \left[\big(M(z)m (z)\big)_f^{f'} \right ]_-= R(z)
 \delta_f^{f'}  \cr
 & \left[\big(m(z) M(z)\big)_{\tilde f}^{\tilde f'} \right]_-
 = R(z) \delta_{\tilde f}^{\tilde f'} \cr
 }}
Here $m(z)$ and $M(z)$ are matrices in flavor space.  We multiply
such matrices as $(AB)^f_{f''}= A^f_{f'} B^{f'}_{f''}$ and $\tr$
denotes a trace over the flavor indices.  Below we will solve
these equations.

We start by solving the third equation in \anoeff.  We write it as
 \eqn\anoeffthi{W'(z) R(z) +{1 \over 4} f(z)  = R(z)^2}
with a polynomial $f(z)=-4\left[W'(z) R(z)\right]_+$ of degree
$n-1$.  Its solution is
 \eqn\solveR{2R(z)= W'(z) -\sqrt{W'(z)^2 +f(z)}.}
This solution is parameterized by the $n$ coefficients in $f(z)$.
We see that $R(z)$ has cuts in the complex $z$ plane.  In the
semiclassical approximation of small $f$, each cut $A_i$ is
naturally associated with a zero of $W'$, $a_i$.

It is natural to analytically continue $z$ through the cuts to a
second sheet.  The double-sheeted complex plane is a Riemann
surface $\Sigma$.  Define
 \eqn\yRrel{y(z)=W'(z) - 2R(z)}
and write \solveR\ as an equation for a Riemann surface $\R$
 \eqn\riems{y^2=W'(z)^2 + f(z).}
This genus $n-1$ Riemann surface was introduced in \CachazoJY. We
can
 naturally understand $\R$ as a double cover of the complex
 $z$-plane branched at the
roots of $W'(z)^2+f(z)$. There are $n$ branch cuts; as above we
denote as $A_i$ a contour that circles around the $i^{th}$ cut.
Most of the functions we will consider in this work have
singularities at infinity; if the points on $\R$ with $z=\infty$
are removed, the $n$ cycles $A_i$ with $i=1,...,n$ are
independent.  Instead of parameterizing $\R$ by the coefficients
in $f$, we alternatively parameterize it by the variables
\eqn\sicoo{ S_i = \tp\oint_{A_i}R(z)dz  \qquad i= 1,..., n.}

Each point on the complex $z$-plane labeled by a given value of
$z$ corresponds to a pair of points on $\R$. To specify a point
$\q$ on $\Sigma$, we must give its $z$ coordinate $z(\q)$ and also
its $y$ coordinate $y(\q)$.  For each point $\q$ on the first
sheet, there is an image point $\tilde \q$ on the second sheet,
with $z(\tilde \q)=z(\q)$ and $y(\tilde \q)=-y(\q)$.  Actually,
the terminology ``first sheet'' and ``second sheet'' is useful
primarily in a semiclassical limit in which  $z$ is large; more
generally we would simply refer to a pair of points $\q$ and
$\tilde \q$ with the same value of $z$. Clearly, we have also
 \eqn\Rztz{\eqalign{
 &W'(\q)=W'(\tilde \q) \cr
  &R(\q) + R(\tilde \q) = W'(z).}}
We will sometimes adopt a convenient though slightly imprecise
notation, writing $z=\q$ when we mean $z=z(\q)$ and $y=y(\q)$, or
$z\to \q$ when we mean $z\to z(\q)$ and $y\to y(\q)$.

The branch of the square root in \solveR\ is chosen such that for
large $z$ in the first sheet we recover the semiclassical answer
 \eqn\semiclz{R(z) \approx - {f(z) \over 4 W'(z)}={S \over z} +
 \CO(1/z^2) .}
(A simple contour deformation argument shows that $S=\sum_i S_i$.)
Then, using \Rztz\ it is easy to see that the asymptotic limit in
the second sheet is
 \eqn\semicltz{R( z) = W'(z)+ \CO(1/ z)=
 g_n z^n + \CO( z^{n-1}).}

\newsec{Solving For $M(z)$ And $T(z)$}

In this section, we consider $R(z)$ of \solveR\ as given and solve
for $M(z)$.  This means that we solve for $M(z)$ as a function of
$S_i$. Moreover, in this section, we will find a solution which
corresponds in the semiclassical limit to the pseudo-confining
vacua \Phiex; in other words, we set all $r_I=0$. For such
solutions, we impose that $M(z)$ and therefore also $T(z)$ are
regular in the first sheet except for the cuts. As we will see,
other solutions correspond to the Higgs vacua \Phiexh. We will
discuss below the question of whether these branches are connected
to the pseudo-confining vacua.

\bigskip\noindent{\it Solving For $M(z)$}

$M(z)$ is solved from the last two equations in \anoeff
 \eqn\openeq{\eqalign{
 &\left[\left(M(z)m(z)\right)_f^{ f'} \right]_-= R(z)
 \delta_f^{ f'}\cr
 &\left[\left(m(z)M(z)\right)_{\tilde f}^{\tilde f}\right]_- =
 R(z) \delta_{\tilde f}^{\tilde f'}\ . }}
We determine the polynomials $[mM]_+$ and $[Mm]_+$ such that
$M(z)$ is regular in the first sheet \SeibergJQ.

We claim that the solution with these boundary conditions is
 \eqn\Msolve{M(z) = -\sum_{i=1}^n \tp\oint_{A_i} {R(x) \over x-z } {1
 \over m(x)} dx}
where $A_i$ are the cuts in $R(z)$.  To prove it, we note that the
integrand in \Msolve\ decays as $\CO({1/x^2})$ at infinity in the
first sheet (with the branch of the square root given in \solveR)
and therefore by contour deformation, \Msolve\ is equivalent to
 \eqn\Msolvea{M(z)=-\sum_{i=1}^n \tp\oint_{A_i} {R(x) \over x-z} {1 \over
 m(x)} dx=R(z) {1 \over m(z)} - \sum_{I=1}^L {R(\q_I) \over z-z_I}
 \tp\oint_{z_I} {1 \over m(x)} dx.}
To obtain this formula, we have used contour deformation on the
first sheet; the term $R(z)/m(z)$ comes from the pole of the
function $1/(x-z)$, and the second term comes from the poles of
matrix elements of $1/m(x)$.  These latter poles are at $z=z_I$,
the points at which $\det\,m=0$.  Each point $z=z_I$ corresponds
to a pair of points $\q_I, \tilde \q_I$ on $\Sigma$, where $\q_I$
is on the first sheet and $\tilde \q_I$ on the second; since the
contour deformation in \Msolvea\ has been done on the first sheet,
the residue at $z=z_I$ involves $R(\q_I)$, the value of $R$ on the
first sheet. To verify that \Msolvea\ satisfies \openeq, we
proceed as follows. Multiplying the right hand side of \Msolvea\
 by $m(z)$ from the right or the left, the first term contributes
$R(z)$ times a unit matrix.  And the second term leads to a
polynomial in $z$ because
 \eqn\propoly{\eqalign{
 {m(z)\over z-z _I} \oint_{z_I} {1 \over m(x)} dx=&{m(z)-m(z_I)
 \over z-z_I} \oint_{z_I} {1 \over m(x)} dx+  {m(z_I)
 \over z-z_I} \oint_{z_I} {1 \over m(x)} dx \cr
  =&{m(z)-m(z_I) \over z-z_I } \oint_{z_I} {1 \over m(x)} dx+  {1
 \over z-z _I} \oint_{z_I} m(x){1 \over m(x)} dx\cr
  =&{m(z)-m(z_I) \over z-z_I } \oint_{z_I} {1 \over m(x)} dx}}
is a polynomial in $z$.  This verifies that our formula for $M(z)$ does
obey \openeq.

Because of the polynomial ambiguity in \openeq, its solution is
not unique. The particular solution \Msolve\ of  was chosen to be
regular at the points $\q_I$, which lie at large values
 $z=z_I$ that (if we are near
a semiclassical limit) are outside the integration contours in
\Msolve. Regularity at $\q_I$ is, as we have seen, the right
behavior for the pseudo-confining solutions with $r_I=0$.
However, the solution \Msolve\ actually has a pole at the points
$\tilde \q_I$ on the second sheet.  This behavior is evident in
\Msolvea: as $R(\tilde \q_I)\not= R(\q_I)$, the singularities of
the two terms on the right hand side of \Msolvea\ do not cancel at
$\tilde \q_I$, though they cancel at $\q_I$.  At first sight, the
meaning of the singularity of $M(z)$ on the second sheet is rather
mysterious, since the second sheet is invisible classically.
  Its interpretation will ultimately
become clear.

\bigskip\noindent{\it Solving For $T(z)$}

Having found $R(z)$ and $M(z)$, we can solve for $T(z)$. We solve
for $T(z)$ as a function of $S_i$. The polynomial ambiguity in the
equations is determined by imposing that $\tp\oint_{A_i} T(z) dz =
N_i$.  Since we use the solution for $M(z)$ given by \Msolve\ or
equivalently by \Msolvea, the result obtained here is relevant to
the pseudo-confining vacua with $r_I=0$.

First we need to calculate $\left[\tr\ m'(z)M(z)\right]_-$ with
$M(z)$ given in \Msolvea.  We get
 \eqn\mpM{\eqalign{
 &\left[\tr\ m'(z)M(z)\right]_-=\left[R(z)\tr\ m'(z){1 \over
 m(z)}\right]_-  - \sum_{I=1}^L \left[{R(\q_I) \over z-z_I}
 \tp\oint_{z_I}\tr\ m'(z) {1 \over m(x)} dx\right]_-\cr
 &\qquad =
 R(z) {B'(z)\over B(z)} - \sum_{I=1}^L {R(\q_I)\over z-z_I}
 - \sum_{I=1}^L \left[R(\q_I)\tr\ {m'(z)-m'(z_I) \over z-z_I}
 \tp\oint_{z_I}{1 \over m(x)} dx\right]_-\cr
 &\qquad ={ B'(z)( W'(z)- y(z))\over 2B(z)} -\sum_{I=1}^L
 {W'(z_I)-y(\q_I)\over 2(z-z_I)}
  \; .}}
We used the relation
$\tr\ m' {1 \over m}= {B'\over B} $ and the
fact that the first two terms in the second line have only
negative powers of $z$ and the third term is polynomial in $z$.

Since $B$ and $W'$ are polynomials and the zeros of $B$ are $z_I$,
 \eqn\useid{{B'(z)W'(z)\over 2B(z)} -\sum_{I=1}^L
 {W'(z_I)\over 2(z-z_I)} =\left[{B'(z)W'(z)\over 2B(z)}\right]_+
 =\left[{B'(z)y(z)\over 2B(z)}\right]_+\ ,}
where the last step
is correct only in the first sheet (on which $W'-y$ vanishes
for large $z$).  Therefore
\mpM\ becomes
 \eqn\anoeqa{\left[\tr\ m'(z)M(z)\right]_-=-\left[{ B'(z)y(z)\over
 2B(z)}\right]_- +\sum_{I=1}^L {y(\q_I)\over 2(z-z_I)}.}

Supersymmetric vacua are described by solutions of \anoeff\ with
$w_\alpha(z)=0$. So the first equation in \anoeff\ reduces
 \eqn\anoeq{\left[W'(z) T(z)\right]_- -2R(z)T(z)+\tr\
 \left[m'(z)M(z)\right]_-=0,}
 and
can be written as
 \eqn\anoeq{\left[y(z) T(z)\right]_- =-\tr\ \left[m'(z)M(z)\right]_-
 =\left[{ B'(z)y(z)\over
 2B(z)}\right]_- -\sum_{I=1}^L {y(\q_I)\over 2(z-z_I)}.}
 Hence
 \eqn\Tsolv{ T(z)={ B'(z)\over
 2B(z)}-\sum_{I=1}^L {y(\q_I)\over 2y(z) (z-z_I)}+ {c(z) \over y(z)}}
where
 \eqn\cpol{c(z)= \left\langle \Tr {W'(z)-W'(\Phi) \over
 z-\Phi}\right\rangle - {1\over 2}  \sum_{I=1}^L {W'(z)-W'(z_I)
 \over z-z_I}}
is a polynomial of degree $n-1$.

For large $z$, the first term in \cpol\ behaves as $Ng_n z^{n-1}$
since $\langle \Tr {\bf 1}\rangle =N$. The second term behaves as
$-{1\over 2}L g_nz^{n-1}$. This is enough to determine the large
$z$  behavior of $T(z)$,
 \eqn\larT{T(z)=\cases{
 {N\over z} + \CO({1\over z^2})& in the first sheet\cr
 {L-N\over   z} + \CO({1\over   z^2})&
 in the second sheet .\cr
 }}
For the pseudo-confining vacua, we expect $T(z)$ to be regular at
the points $\q_I$ on the first sheet. This indeed follows from
\Tsolv, as $y(\q_I)/y(z)\to 1$ for $z\to \q_I$.  On the second
sheet, since $y( \q_I)/y(z)\to -1$ for $z\to \tilde \q_I$, $T(z)$
has a pole, $T(z)\sim 1/(z-z_I)$ for $z\to \q_I$.  Why the residue
of this pole is precisely 1 will be clear when we compare to the
Higgs vacua.

Finally, $c(z)$ can be determined by the requirements
 \eqn\Treq{\tp\oint_{A_i} T(z) dz=N_i\ .}

Let $P$ denote the point $z=\infty$ on the first sheet and $\tilde
P$ the point $z=\infty$ on the second sheet. We can summarize the
singularities of $T$ as follows: $T(z)dz$ has simple poles at $P$,
$\tilde P$, and $\tilde \q_I$ with residues $-N$, $-L+N$ and $1$
respectively.  Elsewhere it is regular.

\newsec{Higgs Vacua And Singularities In The First Sheet}

Consider varying the parameters in $m_f^{\tilde f}(z)$ with fixed
$W(\Phi)$ and fixed $S_i$. The Riemann surface $\Sigma$ is
unchanged but the zeros $z_I$ of $B(z_I)$ change. Let us move one
of them, say $z_1$, through one of the cuts, say $A_1$.  Of the
two points $\q_1$ and $\tilde \q_1$ on $\Sigma$ that correspond to
$z=z_1$, our solutions for $M$ and $T$ have poles at $\tilde
\q_1$, which is on the second sheet, but not at $\q_1$.  When
$z_1$ passes through the cut, the two points $\q_1$ and $\tilde
\q_1$ are exchanged. We end up with solutions for
 $M$ and $T$ which obey the equations but do
not satisfy the boundary conditions that we have imposed so far;
they are singular at $\q_1$.

This process has a simple physical interpretation.  We started
semiclassically with $\langle\Phi \rangle$ whose eigenvalues, as shown
in \Phiex, are approximately equal to the roots $a_i$ of $W'$; this
corresponds to having
all $r_I=0$.  In a semiclassical limit, the
 cut $A_1$ is near $a_1$.  When $z_1$ passes through the cut,
it is near $a_1$ and the solution \Phiex\ is near the solution
\Phiexh\ that describes the Higgs branch.  At this stage, the
strong quantum dynamics are important and a semiclassical
treatment is not precise enough. Passing $z_1$ through the cut and
taking it to be again large (or at least far away from all cuts),
we may find ourselves in a  Higgs branch with $r_1=1$.  On this
branch, $M$ and $T$ are expected to have poles on the first sheet.
Thus, our proposal is that in a process in which $z_1$ moves
through one of the cuts, a branch with $r_1=0$ is continuously
transformed into a branch with $r_1=1$.

Let us see how this occurs in terms of our solution. The analytic
continuation of \Msolvea\ is completely clear: when $z_1$ moves
through the cuts, $\q_1$ and $\tilde \q_1$ are exchanged.  Thus,
\Msolvea\ transforms into
 \eqn\Msolveaa{{M(z)=R(z) {1 \over m(z)} - {R(\tilde \q_1)\over z-z_1}\tp\oint_{z_1}{1\over
m(x)} dx -\sum_{I=2}^L {R(\q_I) \over z-z_I}
 \tp\oint_{z_I} {1 \over m(x)} dx.}}
To derive this from \Msolve, we note that as $z_1$ moves, the cut
$A_1$ must be continuously deformed to avoid $\q_1$ and $\tilde
\q_1$.  By the time $\q_1$ and $\tilde \q_1$ have been exchanged,
$A_1$ has transformed to a contour $A_1'=A_1+C_1-\tilde C_1$,
where $C_1$ and $\tilde C_1$ are small contours that,
respectively, encircle $\q_1$ and $\tilde \q_1$ in the
counterclockwise directions.  \Msolve\ is thus replaced by
\eqn\Msolved{M(z)=-\sum_{i=1}^n\tp\oint_{A_i}{R(x)\over x-z}
{1\over m(x)}dx +\tp\oint_{C_1}{R(x)\over x-z} {1\over m(x)}dx
+\tp\oint_{\tilde C_1}{R(x)\over x-z} {1\over m(x)}dx.} The extra
two terms in \Msolved\ have the effect of cancelling the
singularity at $\tilde \q_1$ and adding a singularity at $\q_1$.

\ifig\oneloopdiag{$(a)$ Original configuration with $q_1$ on the
upper sheet (black dot) enclosed by the $C_1$ contour and $\tilde
q_1$ on the lower sheet (white dot) enclosed by the $\tilde C_1$
contour. Also shown is the first cut (wide black line) enclosed by
the $A_1$ contour. $(b)$ Same configuration as in $(a)$ after
$\q_1$ and $\tilde \q_1$ are passed through the first cut. The
$A_1$ contour is deformed as shown. A new $A'_1$ cycle that only
encloses the cut is then given by $A'_1 = A_1 + C_1 - \tilde C_1$}
{\epsfxsize=0.6\hsize\epsfbox{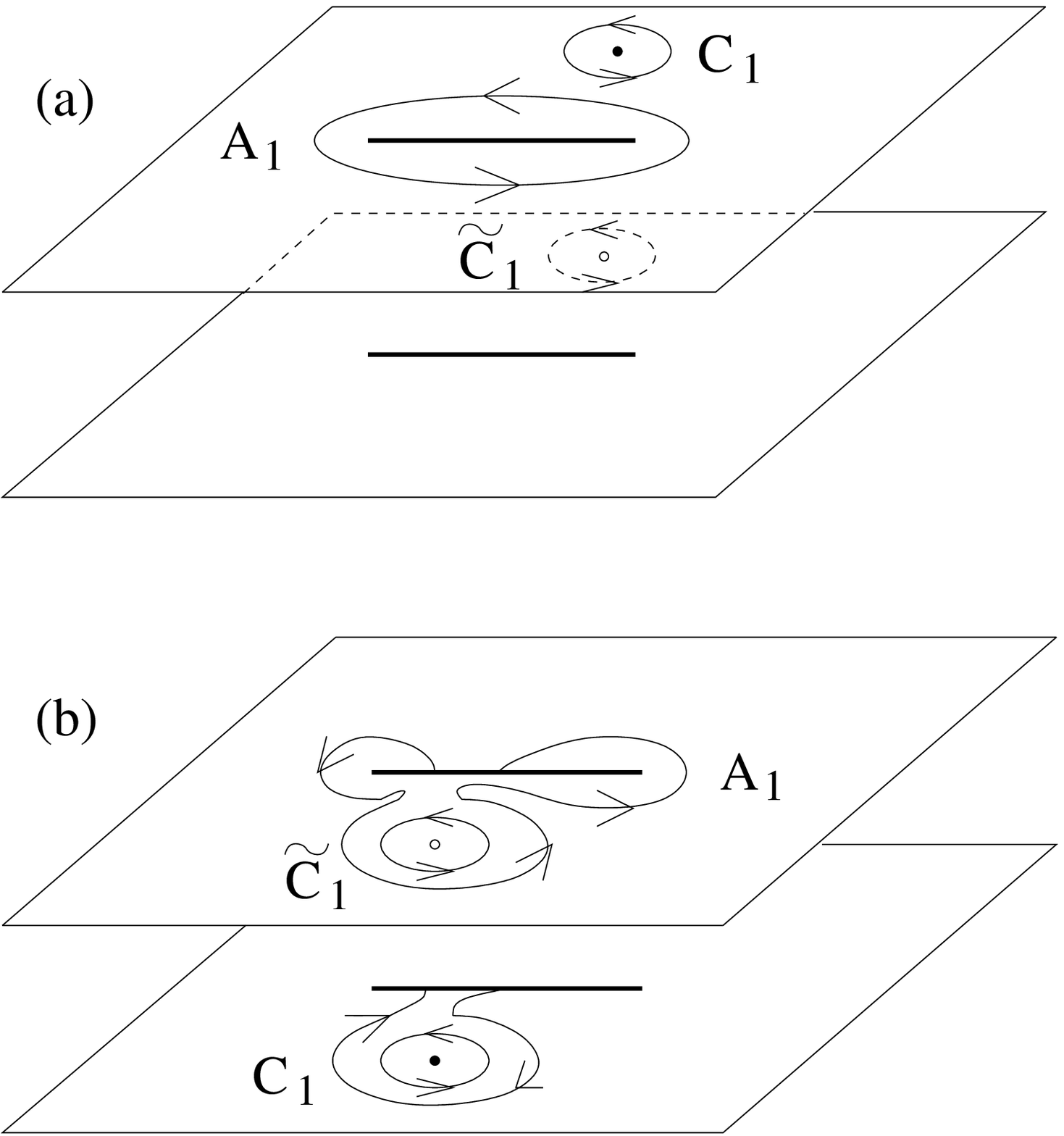}}

Similarly, the singularity of $T(z)$ in the second sheet at
$\tilde \q_1$ moves to the first sheet to $\q_1$. Thus, after
$z_1$ passes through the $A_1$ cut, we have $\tp\oint_{\tilde
C_1}T(z)dz=0$, $\tp\oint_{C_1}T(z)dz=1$.  In the course of the
transition, the integers $N_i=\tp\oint_{A_i}T(z)dz$ cannot change.
After the transition, $A_1$ has transformed to
$A_1'=A_1+C_1-\tilde C_1$, so after the transition
$N_1=\tp\oint_{A_1'}T(z)dz$ and hence
$N_1-1=\tp\oint_{A_1}T(z)dz$. Thus, the effective value of $N_1$
has been reduced by 1 in this transition.

More generally, for arbitrary $r_I=0,1$, the solution for $M$
becomes
 \eqn\Mstrg{\eqalign{
 M(z) = &R(z) {1\over m(z)} - \sum_{I=1}^L (1-r_I){R(\q_I)\over
 z- z_I } \tp\oint _{z_I} {1 \over m(x)} dx -\sum_{I=1}^L r_I{R(
 \tilde \q_I)\over z- z_I } \tp\oint _{z_I} {1 \over m(x)} dx\cr
 =&R(z) {1\over m(z)} - \sum_{I=1}^L {r_IW'(z_I) +(1-2r_I)
 R(\q_I) \over z- z_I } \tp\oint _{z_I} {1 \over m(x)} dx}}
where we used \Rztz\ to express $R(\tilde \q_I)=W'(z_I)-R(\q_I)$.
We would like to make a few comments about this expression:
 \item{1.} $M(z)$ has a singularity with residue $r_I(2 R(\q_I)
- W'(z_I)) \tp\oint _{z_I} {1 \over m(x)} dx$ at $\q_I$ and a
singularity with residue $(r_I-1)(2 R(\q_I) - W'(z_I)) \tp\oint
_{z_I} {1 \over m(x)} dx$ at $\tilde \q_I$.  In particular,
singularities are at $\tilde \q_I$ if $r_I=0$ and at $\q_I$ if
$r_I=1$. \item{2.} In the classical limit of zero $R(z)$, this
expression coincides with our classical answer \resMc.  Note that
the residue of $M$ at $\q_I$ is $r_I(2 R(\q_I) - W'(z_I)) \tp\oint
_{z_I} {1 \over m(x)} dx$, which is corrected from the classical
answer $-r_IW'(z_I) \tp\oint _{z_I} {1 \over m(x)} dx$.
 \item{3.} It is easy to check, as in \propoly, that the terms
 which are summed over $I$ in \Mstrg\ have the property that when
 they are multiplied by $m(z)$ they lead to a polynomial in $z$.
 These terms change the poles in $M$ and modify
 the polynomials $[mM]_+$ and $[Mm]_+$, but
 do not affect the anomaly equations \openeq.

By repeating the procedure that was used to find \Tsolv, or by simply
analytically continuing \Tsolv\ to let the $z_I$ pass through the
cuts,  we find that the general solution for $T$ is
 \eqn\Tsolvr{ T(z)={ B'(z)\over
 2B(z)}-\sum_{I=1}^L {(1-2r_I) y(\q_I)\over 2y(z) (z-z_I)}+
 {c(z) \over y(z)}.}
We have used the fact that when $z_I$ passes through a cut and
$r_I$ becomes 1, $y(\q_I)$ is transformed to
$-y(\q_I)=(1-2r_I)y(\q_I).$ As expected, for $r_I=0$ the function
is regular at $\q_I$ and has a pole with residue one at $\tilde
\q_I$, while for $r_I=1$ it is regular at $\tilde \q_I$ and has a
pole with residue one at $\q_I$. The number of singularities in
the first sheet is $\sum_I r_I$ and satisfies $L\ge \sum_I r_I \ge
0$ in agreement with the classical reasoning in \sumrI\NLine.

By interpreting the analytic continuation of the pseudo-confining
vacua as $z_I$ pass through cuts as physical Higgs vacua, we can
shed light on many otherwise mysterious features of the solutions
found in section 3 for the pseudo-confining vacua.  For example,
the singularities on the second sheet now have a rationale; upon
analytic continuation, they become semiclassical singularities on
the first sheet in Higgs vacua.  It is also now clear why the
residues of poles of $T(z) dz$ on the second sheet are precisely
1: this is needed to agree with the semiclassical behavior in
Higgs vacua.

All of our analysis has been carried out at fixed $S_i$.  This
corresponds to treating the field theory off-shell; an on-shell
analysis, by contrast, would involve extremizing the
superpotential as a function of the $S_i$.  In effect, in our
analysis, we have ignored the fact that the $S_i$ will change when
the $z_I$ are varied.  In this off-shell analysis, we have found
that $N_i$ can be reduced, for any given $i$, by passing the $z_I$
through the $i^{th}$ cut $A_i$.  Clearly, there must be a limit to
this process, for if it is performed too many times, $N_i$ will
become negative.  Intuitively, we expect that if the $N_i$ are
large enough then the back-reaction on the $S_i$ in moving a
single $z_I$ through a cut is small.  How far can one expect to go
before the back-reaction becomes significant?  We conjecture but
will not try to prove in this paper that a given $N_i$ can be
reduced all the way to $1$ by passing $N_i-1$ of the $z_I$ through
the cut $A_i$.  One cannot expect to go farther, since if $N_i$
changes from 1 to 0, this would change the number of $U(1)$'s in
the low energy gauge group, so such a process cannot occur
continuously. It must be that if one tries to transform $N_i$ all
the way to zero, the cut $A_i$ will close up at the last step. We
return briefly to this issue in section 8.

\newsec{ $T(z)dz$ As A Differential And The Complete Solution}

By definition, $T(z)dz$ is a meromorphic differential on the
Riemann surface $\R$ of genus $n-1$ that was given in  \riems:
\eqn\rsuhh{ y^2=W'(z)^2+f(z).}
$T(z)dz$ has simple poles at $P$, $\tilde P$ and $\tilde \q_I$
with residues $-N$, $N-L$ and $1$ respectively.  We are now going
to introduce some formalism for describing $T$ in more detail.
This will eventually be used in section 7 to prove that on-shell,
the periods of $T$ are all integers.  At the end of the present
section, we will assume this statement and deduce some
consequences.

Ignoring the punctures of $\R$ at $P $ and $\tilde P$, a complete
basis of independent $A$-cycles of $\R$
are the  $A_i$ for $i=1,...,n-1$. If $p_1$ and $p_2$ are any chosed points
in $\R$, then the Riemann-Roch or index theorem can be used to show that
meromorphic differentials on $\R$ with poles at $p_1$ and $p_2$ and no
other singularities do exist.  By normalizing such a differential, we can
make the residues at $p_1$ and $p_2$ precisely $-1$ and $1$; by adding
a holomorphic differential, we can in a unique fashion
make the $A_i$ periods vanish.
Hence, there exists
a unique meromorphic differential $\t_{p_1,p_2}$ whose only
singularities on $\R$ are simple poles at $p_1$ and $p_2$ with
residues $-1$ and $1$ respectively and $\oint_{A_i}\t_{p_1,p_2} =
0$.

Since $T(z)dz$ is singular only at $\tilde \q_I$, $P$ and $\tilde
P$, the differential
\eqn\tmerok{T(z)dz -\left( N \t_{P,\tilde P} + \sum_{I=1}^{L}
\t_{\tilde P,{\tilde \q}_I} \right)   }
is a holomorphic one-form on $\R$. Using a basis for holomorphic
differentials $\{ \h_1,\ldots \h_{n-1}\}$ on $\R$ which is
characterized by $\oint_{A_i}\h_k = \delta_{ik}$, we have
\eqn\holk{T(z)dz = N \t_{P,\tilde P} + \sum_{I=1}^{L} \t_{\tilde P
,{\tilde \q}_I} + \sum_{j=1}^{n-1}h_j \h_j\ .}
The constants $h_i$ are determined from \Treq\ for $i=1,\ldots ,
n-1$ as $h_j=2\pi i N_j$. The condition on $A_n$ is automatically
satisfied. To see this note that on the first sheet the integral
around $A_n$ can be deformed to an integral around $P$ and
$-A_1-\ldots -A_{n-1}$. This leads to $\tp\oint_{A_n} T(z)dz
=N-N_1-\ldots -N_{n-1}= N_n$. We conclude that
\eqn\tfin{T(z)dz =  N \t_{P,\tilde P} + \sum_{I=1}^{L} \t_{\tilde
P,{\tilde \q}_I} + 2\pi i\sum_{j=1}^{n-1}N_j \h_j\ . }

We now consider the noncompact cycles $\{ \B_1,\ldots ,\B_n\}$
with the intersection pairing $A_i \cap \B_j = \delta_{ij}$, $\B_i
\cap \B_j = 0$ (see figure 2).  It is convenient to use the basis
$\{ \b_1 ,\ldots ,\b_{n-1}, \B_n \}$ defined by $\b_i = \B_i -
\B_n$. Clearly, $\b_i$ are compact cycles and together with $A_i$
for $i=1,\ldots ,n-1$ form a canonical basis of $H_1(\R )$.

\ifig\oneloopdiag{Riemann Surface $\R$ for $n=3$ as a double cover
of the $z$ complex plane. Solid lines are on the first sheet while
dashed lines are on the second sheet. Gray circles represent the
location of the branching points of \rsuhh. The $i^{th}$ branch
cut runs between the points $a_i^-$ and $a_i^+$ which collapse to
the $i^{th}$ root, $a_i$, of $W'(z)$ when $f(z)$ is taken to zero.
$\{A_1,A_2,A_3,\B_1,\B_2,\B_3\}$ form a basis of one cycles, on
the surface with punctures, with canonical intersection pairing.
$\{A_1,A_2,\b_1,\b_2,\B_3\}$ is a different choice of basis. $\{
A_1,A_2,\b_1,\b_2\}$ is the canonical basis of one cycles for a
compact Riemann surface of genus $2$. The points $P$ and $\tilde
P$ with coordinates $z(P)=z(\tilde P )=\infty$ and $y(P)=-y(\tilde
P)$ are depicted as a black dot and a white dot respectively.}
{\epsfxsize=0.6\hsize\epsfbox{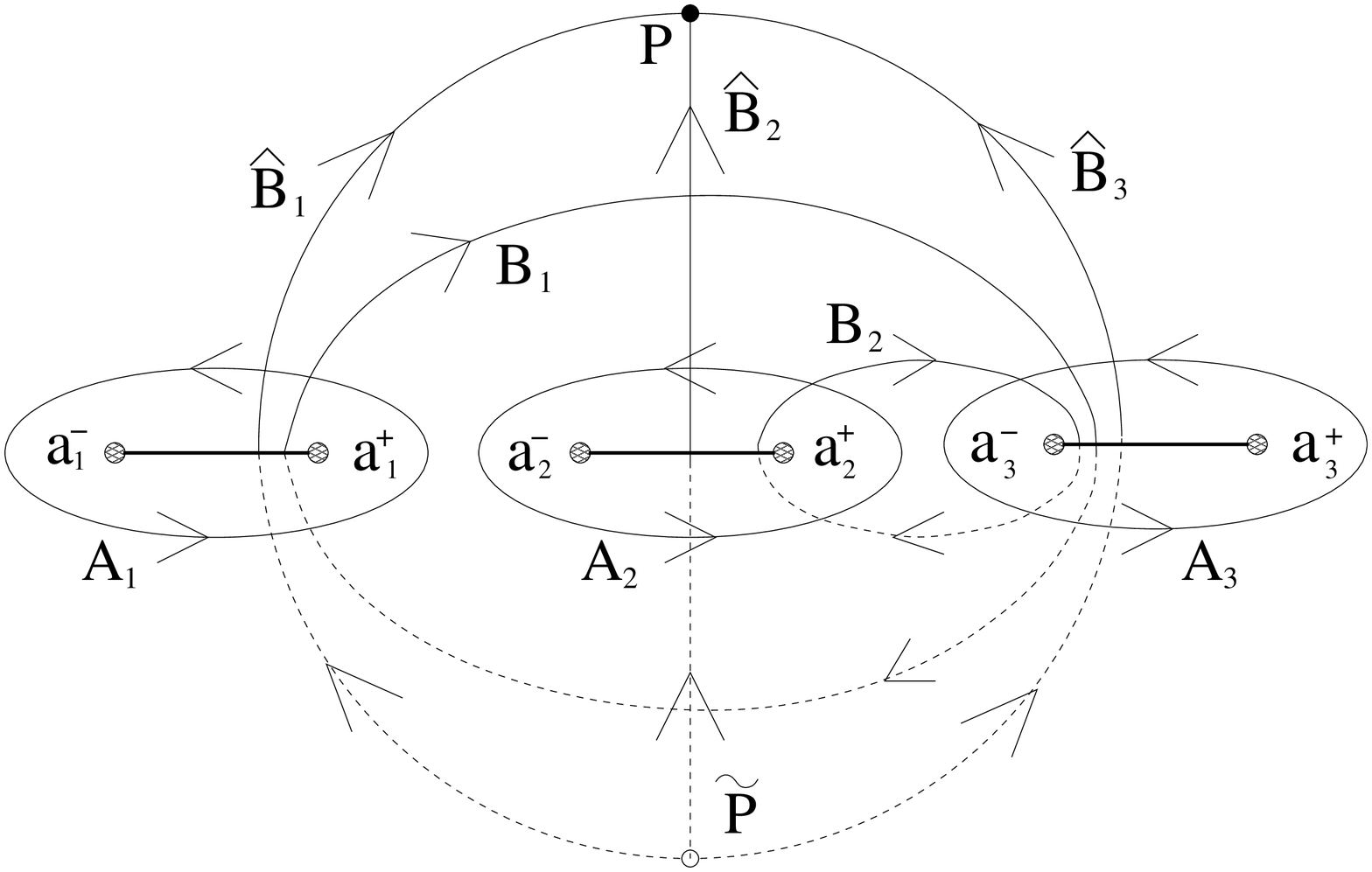}}

Consider first the period
\eqn\perbb{ \oint_{\b_k}T(z)dz =N\oint_{\b_k} \t_{P,\tilde P}
+\sum_{I=1}^{L}\oint_{\b_k}\t_{\tilde P,{\tilde \q}_I}+ 2\pi
i\sum_{j=1}^{n-1}N_j\oint_{\b_k} \h_j.  }
Using the Riemann bilinear relations, we have that for any
$\t_{p_1,p_2}$ (see appendix B for a review),
\eqn\billl{\tp\oint_{\b_k}\t_{p_1,p_2} = \int_{p_2}^{p_1}\h_k}
where the contour in the right hand side is as follows. If $p_1$
and $p_2$ are on the same sheet, the contour is on that sheet. If
$p_1$ and $p_2$ are on different sheets the contour crosses
through cut number $n$. In both cases winding around the $A_k$
cycles can lead to an integer ambiguity. This ambiguity
corresponds to the fact that in order to define the differentials
$\t_{p_1,p_2}$ and to derive the Riemann bilinear relations a
particular choice of representative of one cycles has been made. A
different choice leads to an answer which differs by an integer.
See appendix A for a more detailed discussion of this issue.

Using \billl\ in the first and second terms of \perbb\ and the
fact that $\oint_{\b_k} \h_j = \oint_{\b_j}\h_k$ in the last term
of \perbb\ we get,
\eqn\ghfin{  \tp\oint_{\b_k}T(z)dz =N\int_{\tilde P}^{P}\h_k +
\sum_{I=1}^{L}\int_{{\tilde \q}_I}^{\tilde P}\h_k +
\sum_{j=1}^{n-1}N_j\oint_{\b_j} \h_k. }
Since the first and the last term of \ghfin\ can be combined
\eqn\combj{N\int_{\tilde
P}^{P}\h_k+\sum_{j=1}^{n-1}N_j\oint_{\b_j} \h_k
=N\int_{\B_n}\h_k+\sum_{j=1}^{n-1}N_j\oint_{\b_j} \h_k
=\sum_{j=1}^{n}N_j\int_{\B_j} \h_k,}
\ghfin\ can be written as
\eqn\gfink{  \tp\oint_{\b_k}T(z)dz = \sum_{j=1}^{n}N_j\int_{\B_j}
\h_k + \sum_{I=1}^{L}\int_{{\tilde \q}_I}^{\tilde P}\h_k.}

Finally, we can compute the period of $T(z)dz$ on the non-compact
cycle $\B_n$. This period is infinite and requires a
regularization. We will introduce regularized contours $\rb_i$, as
contours from $\tilde\Lambda_0$ to $\Lambda_0$ passing through the
$i^{th}$ cut. $\Lambda_0$ is a complex number with sufficiently
large $|\Lambda_0|$. $\Lambda_0$ and $\tilde\Lambda_0$ are the
points in the first and the second sheet which correspond to the
same value of $\Lambda_0$.  Only the behavior up to
$\CO(1/\Lambda_0)$ will be relevant for us in the computations.

Using \tfin, we have
\eqn\nonpp{ \int_{\rb_n}T(z)dz = \int_{\rb_n}\left( N\t_{P,\tilde
P} + \sum_{I=1}^L\t_{{\tilde P},{\tilde \q}_I}  + 2\pi
i\sum_{i=1}^{n-1}N_i\h_i\right). }
We want to write it as an integral of $\t_{P,\tilde P}$ over
various cycles.  In the last term we use
$\int_{\rb_n}\h_i=\int_{\tilde
\Lambda_0}^{\Lambda_0}\h_i=\tp\oint_{\b_i}\t_{\Lambda_0,\tilde
\Lambda_0} $. In the second term we use the Riemann bilinear
relations
 \eqn\riebi{\int_S^R\t_{TQ}= \int_Q^T\t_{RS}}
(recall that here and in the following expressions with line
integrals there is an ambiguity in $2\pi i$ times an integer which
is discussed in appendix A) to write
 \eqn\auxexp{\int^{\Lambda_0}_{\tilde \Lambda_0} \t_{{\tilde P}
 ,{\tilde \q}_I}=\int_{{\tilde \q}_I}^{\tilde P}\t_{\Lambda_0,
 \tilde \Lambda_0}=\int_{{\tilde \q}_I}^{\tilde P}\t_{P,
 \tilde \Lambda_0} + \CO(1/\Lambda_0).}
Finally, we  note that
 \eqn\useex{\int_{{\tilde \q}_I}^{\tilde P}\t_{P,
 \tilde \Lambda_0}=\int_{{\tilde \q}_I}^{\tilde \Lambda_0}\t_{P,
 \tilde P} + \pi i  +\CO(1/ \Lambda_0).}
See appendix A for a proof of this result.

Combining these results, \nonpp\ can be written as,
\eqn\ontu{ \int_{\rb_n}T(z)dz = N \int_{\tilde
\Lambda_0}^{\Lambda_0}\t_{P,\tilde P} + \pi i L
+\sum_{i=1}^{n-1}N_i \oint_{\b_i}\t_{P,\tilde P}+ \sum_{I=1}^L
\int_{{\tilde \q}_I}^{\tilde \Lambda_0}\t_{P,\tilde P}
+\CO(1/\Lambda_0)\ . }
where we have used that $\oint_{\b_i}\t_{\Lambda_0,\tilde
\Lambda_0} = \oint_{\b_i}\t_{P,\tilde P} + {\cal O}(1/\Lambda_0)$.

This can be brought to the same form as \gfink\ by using that
$\b_i =\rb_i-\rb_n$,
\eqn\fing{ \int_{\rb_n}T(z)dz
=\sum_{j=1}^{n}N_j\int_{\rb_j}\t_{P,\tilde P} +
\sum_{I=1}^{L}\int_{{\tilde \q}_I}^{\tilde \Lambda_0}\t_{P,\tilde
P} + \pi i L +\CO(1/\Lambda_0)\ . }

Equations \gfink\ and \fing\ for the periods of $T(z)dz$ will be
shown to be related to the field equations coming from the field
theory effective superpotential in  section 7. The field
equations imply that $\tp\oint_{\b_i}T(z)dz$ are integers. Here we
will assume this fact and show that it is enough to determine all
the remaining parameters.

Assuming that $\tp\oint_{\b_i}T(z)dz$ is an integer for all
$i=1,...,n-1$ and recalling \Treq, it follows that $T(z)dz$  has
integer periods over all compact cycles of $\R$. Therefore
$T(z)dz$ has to be a logarithmic derivative of a meromorphic
function on $\R$: $T(z)dz= d\log \psi(z)$ for some $\psi(z)$.
$\psi$ should have a pole at $P$ of order $N$ and zeros at $\tilde
P$ and $\tilde \q_I$ of order $N-L$ and $1$ respectively. For $L
\leq 2N$ such a function can be constructed as follows \NaculichHR
\eqn\fund{ \psi(z)= P(z) + \sqrt{ P^2(z) - \alpha B(z)},}
where the degree $N$ polynomial $P(z)$ and the constant $\alpha$
must be determined.

The condition that \fund\ be single valued on $\R$ is
 \eqn\sigval{ \eqalign{
 &P^2(z) - \alpha B(z)=F(z) H^2(z)  \cr
 &W'(z)^2+f(z) = F(z)Q^2(z),\cr}}
where $Q(z)$, $H(z)$ and $F(z)$ are polynomials in $z$.  Indeed,
these conditions imply that $\sqrt{P^2-\alpha B}=yH(z)/Q(z)$.

The degree of $Q(z)$ corresponds to the number of $N_i$'s that are
taken to be zero. When a given $N_i$ is zero, the corresponding
branch cut in $y(z)$ closes up.

Special cases are the following. \item{1.} deg$Q(z)=0$: This
corresponds to the case with all $N_i\neq 0$ and the deg$W'(z)
\leq N$. \item{2.} deg$H(z)=0$: This corresponds to the case when
the group is completely broken  to $U(1)^N$ and deg$W'(z)\geq N$.

Using that
\eqn\tbe{ T(z)dz = \partial_z \chi dz; \qquad\qquad \psi (z)
=e^{\chi(z)}\  . }
we can explicitly compute the period of $T(z)dz$ on $\rb_n$,
\eqn\diver{  \int_{\rb_n}T(z)dz = \log\left( {4\over \alpha B_L}
\Lambda_0^{2N-L} \right) + {\cal O}\left( {1\over
\Lambda_0}\right)}
where $B_L$ is defined such that $B(z)= B_L \prod (z-z_I)$. This
result is consistent with the asymptotic behavior of $T(z)$ given
in \larT.

In order to determine $\alpha$, note that $B_L$ has mass dimension
${N_f-L}$, so $\alpha$ must have dimension ${2N-N_f}$. The only
parameter with mass dimension that does not depend on the details
of the superpotential is the scale $\Lambda$ of the theory. So
$\alpha$ is equal to $\Lambda^{2N-N_f}$ up to a numerical
constant. This constant depends on the subtraction scheme. We
choose a scheme where the standard $\N=2$ result for the curve
\sigval\ is recovered when $m(z)_f^{\tilde f}= \sqrt{2}(z+m_f)
\delta_f^{\tilde f}$. Note that here $m_f$ is not the canonically
normalized mass. This choice leads to $\alpha/4=
2^{-N_f/2}\Lambda^{2N-N_f}$. It is convenient to define ${\hat
B}(z) =2^{-N_f/2}B(z)$ such that the final result for $T(z)$ is
given by,
\eqn\conk{ \left\langle \Tr {1\over z-\Phi }\right\rangle =
{\del\over \del z} \log \left( P(z) + \sqrt{P^2(z) -
4\Lambda^{2N-N_f} {\hat B}(z)} \right).}

Through equation \sigval\ we can now determine also $f(z)$ and
therefore $y(z)$ and $R(z)$.  This means that all the observables
of the theory are now determined.  Of course, we still need to
justify the claim that the compact periods of $T$ are all
integers.  We postpone this to section 7.

\newsec{Matrix Model Computations}

In this section, we will explicitly compute the effective
superpotential from the matrix model.

Following \SeibergJQ, only the sphere and disk contribution to the
free energy of the corresponding matrix model are relevant. Let us
denote them $\CF_0$ and $\CF_1$ respectively. Explicitly, we have
to compute
\eqn\mmin{  \exp \left( \! -{{\hat N}^2\over S^2}\CF_0 - {\hat N
\over S}\CF_1 +\dots \! \right) \! = {1\over {\rm vol}\; U(\hat N)
}\int \! {d \hat\Phi \over \Lambda^{{\hat N}^2}} {d {\hat Q} d
{\hat{\tilde Q}} \over \nu^{\hat N N_f}}\exp \left( \! -{{\hat
N}\over S}\left[ \Tr W(\hat\Phi ) + {\hat{\tilde Q}} m( {\hat\Phi}
) {\hat Q}\right] \right)}
where $\Lambda$ is a dimension one constant which will later be
identified with the dynamically generated scale of the field
theory. $\nu$ is defined such that if all quarks are taken to have
masses equal to $\Lambda$, i.e.\ $m(\hat\Phi )_f^{\tilde
f}=\Lambda\delta_f^{\tilde f}$, then the matrix integral over them
is equal to one; i.e.
\eqn\nueq{  \nu =\int dx \,d {\bar x} \,\,e^{-{\hat N} \Lambda
x{\bar x} / S} = {\pi S\over {\hat N}\Lambda} = \exp\left( -\log
{\Lambda \hat N \over \pi S}\right)}
where $x$ and ${\bar x}$ are two integration variables\foot{ We
can also redefine $\hat Q $ and $\tilde {\hat Q}$ such that
 \eqn\mmina{  \exp \left( -{{\hat N}^2\over S^2}\CF_0 - {\hat
N \over S}\CF_1 +\dots\right) = {1\over {\rm vol}\; U(\hat N)
}\int {d \hat\Phi \over \Lambda^{{\hat N}^2}} {d {\hat Q} d
{\hat{\tilde Q}} \over \nu^{\hat N N_f}}\exp \left( -{{\hat
N}\over S}\Tr W(\hat\Phi ) - {\hat{\tilde Q}} m( {\hat\Phi} )
{\hat Q} \right).}
and $\nu=\pi/\Lambda$.}.

For large $\hat N$ the volume of $U(\hat N)$ is
\refs{\Macdonald,\OoguriGX},
\eqn\volun{ {\rm vol}\; U(\hat N) = \exp\left( - {{\hat N}^2\over
2} \left[ \log {\hat N\over 2\pi } - {3\over 2}\right]+ \CO(1)
\right).}
Therefore,
\eqn\grkw{ {1\over \Lambda^{{\hat N}^2} {\rm vol}\; U(\hat N)}  =
\exp\left( -{{\hat N}^2\over S^2} \left[ \half S^2 \left(\log
{2\pi\Lambda^2 \over \hat N}+{3\over 2}\right)\right] + \CO(
1)\right). }

At weak coupling the potential in \mmin\ has several minima.  We
will focus on an expansion around minima where $\hat N_i$ of the
eigenvalues of $\hat \Phi$ are at $a_i$ and $\hat Q=\hat{\tilde
Q}=0$. There are also minima with $r_I=0,1$ eigenvalues of $\hat
\Phi$ at $z_I$ and nonzero $\hat Q$, $\hat{\tilde Q}$.

The integral over $\hat Q$ and $\hat{\tilde Q}$ can be explicitly
carried out for diagonal $\hat\Phi$ with eigenvalues $\lambda_k$,
\eqn\qint{ \int d {\hat Q} d {\hat{\tilde Q}} \exp \left( -{{\hat
N}\over S} {\hat{\tilde Q}} m( {\hat\Phi} ) {\hat Q} \right)  =
\exp\left( -{\hat N\over S} \left[ N_f S\log {\hat N\over \pi S} +
{S\over \hat N}\sum_{k=1}^{\hat N}\log B(\lambda_k ) \right]
\right).}
Recall that $B(z)$ has been defined in \Bdef\ to be equal to the
determinant of $m(z)$.

Reducing the integration over $\hat\Phi$ to its eigenvalues
produces the Vandermonde determinant. This implies that the only
integration left in \mmin\ can be written as
\eqn\effmm{ \lim_{{\hat N}\to \infty } \int \prod_{k=1}^{\hat
N}d\lambda_k \exp\left( -{\hat N\over S}\sum_{k=1}^{\hat
N}W(\lambda_k ) + 2 \sum_{i<j}\log(\lambda_i - \lambda_j) -
\sum_{k=1}^{\hat N}\log B(\lambda_k ) \right) .}
This integral can be evaluated using a saddle point approximation
and taking into account the measure factors gives
\eqn\sadd{ \exp\left( - {{\hat N}^2\over S^2}\left[ \int d\lambda
\rho(\lambda)W(\lambda ) -\int\!\!\int d\lambda d\lambda'
\rho(\lambda )\rho(\lambda' )\log\left|{\lambda-\lambda' \over
\Lambda} \right| \right] - {\hat N\over S}\left[\int d\lambda
\rho(\lambda )\log {B(\lambda)\over \Lambda^{N_f}} \right]\right)
}
where $\rho(\lambda )$ is the eigenvalue distribution normalized
to $\int \rho(\lambda) d\lambda=S$. This density function gives
$\hat N_i$ eigenvalues in the $i^{th}$ cut.

{}From \sadd, $\CF_0$ and $\CF_1$ can be identified. For $\CF_0$
we have
\eqn\fzer{ \CF_0 = \int d\lambda \rho(\lambda)W(\lambda )
-\int\!\!\int d\lambda d\lambda' \rho(\lambda )\rho(\lambda'
)\log\left|{\lambda-\lambda'\over \Lambda }\right|. }
As a check of the choice of measure (similar to computations by
Dijkgraaf and Vafa), we can take a superpotential of degree two,
i.e.\ $W =\half m \Tr \hat\Phi^2$ and $m(\hat\Phi)_f^{\tilde f} =
\Lambda \delta_f^{\tilde f}$. Explicit computation of the matrix
integral \mmin\ and \fzer\ can be done. Both computations give the
same answer,
\eqn\masst{ \CF_0 = {1\over 2}S^2 \log {m\Lambda^2\over S}
+{3\over 4}S^2.}

For $\CF_1$ we have,
\eqn\fone{ \CF_1 = \int d\lambda \rho(\lambda )\log
{B(\lambda)\over \Lambda^{N_f}}.}

Let us explicitly compute the disk amplitude in terms of the
resolvent\foot{Our definition of the resolvent differs by a sign
from the standard matrix model convention (see for example
\DiFrancescoNW\ ) which is given by ${1\over {\hat N}}\Tr {1\over
\hat\Phi -z}$. } $R(z)$. The integral in \fone\ can be written as
a contour integral by recalling that $\rho(\lambda ) =-{1\over
2\pi i} {\rm disc}\; R(z)|_{z=\lambda }$,
\eqn\cohh{\int d\lambda  \rho(\lambda )\log B(\lambda) =
\tp\sum_{i=1}^n \oint_{A_i} dz R(z)\log B(z).}
In the logarithm we take the principal branch $\log r = \log |r| +
i \alpha$ with $-\pi < \alpha \le \pi$. Since $B(z)= B_L
\prod_{I=1}^L (z-z_I)$ it has branch cuts which we take to run
between $z_I$ and infinity ($P$ in the upper sheet and $\tilde P$
in the lower sheet). We make sure that they do not intersect the
cuts of $R(z)$, such that \cohh\ has no ambiguity.

The integral \cohh\ can be written as
\eqn\sigd{\tp\sum_{i=1}^n \oint_{A_i}dz R(z)\log B(z) = S \log B_L
+ \tp\sum_{I=1}^L\sum_{i=1}^n \oint_{A_i}dz R(z)\log (z-z_I).}
Now we deform the contour integral.  We can do it either in the
upper or the lower sheet.  The simplest possibility is with $R(z)$
in the upper sheet because at infinity $R = S/z+{\cal O}(z^{-2})$.
The contour integral is given by a sum of two terms. The first is
an integral around infinity: a circle of radius $|\Lambda_0|$.
This gives
\eqn\ginn{{S L\over 2\pi i} \int_{\Lambda_0}^{e^{2\pi i}
\Lambda_0} {\log z \over z} dz = S L \log(- \Lambda_0).}
The second contribution is a sum of contour integrals around the
cuts of $\log (z-z_I)$ which run from $\q_I$ to $\Lambda_0$:
\eqn\insec{\tp\sum_I \int_{C_I} dz R(z) \log (z-z_I) = - \sum_I
\int_{\q_I}^{\Lambda_0} R(z) dz.}
The logarithmic divergences of the two contributions \ginn\ and
\insec\ cancel.  This is not surprising, because the original
contour integral \cohh\ around the cuts of $R$ is finite.

We conclude that the disk amplitude is
\eqn\sucf{ \CF_1 = S \log\left({(-\Lambda_0)^LB_L\over
\Lambda^{N_f}}\right) - \sum_I \int_{\q_I}^{\Lambda_0} R(z) dz.}
We can also use $R(z)=\half (W'(z)-y(z))$ to express the integral
as an integral of $y$, and then we can write it as an integral in
the second sheet
 \eqn\sucfs{ \CF_1 = - \half \sum_I
\int_{\tilde \q_I}^{\tilde \Lambda_0}y \,dz +  S
\log\left({(-\Lambda_0)^LB_L\over \Lambda^{N_f}}\right) -\half L
W(\Lambda_0) +\half \sum_I W(z_I) .}

\vskip 5cm
\bigskip\noindent{\it Matrix Model Determination Of $M(z)$}

Before concluding this section, let us present a matrix model
determination of \Msolve. Recall the identification made in
\SeibergJQ,
\eqn\idenJ{\langle {\hat M}(z) \rangle = \langle M(z) \rangle ,}
where $\langle M(z)\rangle$ is the field theory expectation value
of the operator defined in \chiralopf\ and $\langle \hat
M(z)\rangle$ is the average of the matrix function ${\hat{\tilde
Q}}{1\over z-\hat\Phi}\hat Q$. In order to compute the latter, we
can first perform the gaussian integration over the matter fields
$\hat{\tilde Q}$ and $\hat Q$ in \mmin\ for a given eigenvalue of
$\hat\Phi$. This leads to the insertion of ${1\over \lambda_k -
z}{1\over m(\lambda_k)}$ in \effmm. In the saddle point
approximation, we get $\langle {\hat M}(z) \rangle$ by integrating
the single eigenvalue result over the eigenvalue distribution,
i.e.,
\eqn\diseg{\langle {\hat M}(z) \rangle = \int d\lambda {\rho
(\lambda )\over \lambda - z}{1\over m(\lambda)}. }

Using that $\rho(\lambda) = -{1 \over 2\pi i }{\rm
disc}R(z)|_{z=\lambda} $ \diseg\ can be written as,
\eqn\almt{\langle {\hat M}(z) \rangle = -
\tp\sum_{i=1}^n\oint_{A_i} {R(x)\over x-z}{1\over m(x)}dx}
which agrees with \Msolve.

\newsec{The Off-Shell Effective Superpotential}

The field theory effective superpotential receives contributions
from the sphere, $\CF_0$, as well as from the disk, $\CF_1$,
amplitudes in the matrix model \SeibergJQ.

The sphere contribution up to terms with two derivatives of $S_i$
is
\eqn\suef{W_{eff}|_{\rm sphere} = \sum_{i=1}^n N_i {\del
\CF_0\over \del S_i}+ 2\pi i\sum_{i=1}^{n-1} b_i S_i.}
Note that no explicit $\tau^{ft}_0 S$ appears in \suef, where
$\tau^{ft}_0$ is the bare coupling of the field theory. This is
due to the way the matrix integral \mmin\ is normalized. In this
normalization, $\CF_0$ is a finite quantity.

A detailed computation of ${\del \CF_0\over \del S_i}$ can be
found in appendix B. The result is given by
\eqn\appr{{\del \CF_0\over \del S_i}= -\half\int_{\rb_i}y(z)dz +
W(\Lambda_0) - 2S \log\left(-{\Lambda_0\over \Lambda } \right) +
{\cal O}(1/\Lambda_0). }
Therefore \suef\ is, up to ${\cal O}(1/\Lambda_0)$ terms,
\eqn\atlast{ W_{eff}|_{\rm sphere } = -\half \sum_{i=1}^n
N_i\int_{\rb_i}y(z)dz + N W(\Lambda_0) - 2N S
\log\left(-{\Lambda_0\over \Lambda } \right)+ 2\pi
i\sum_{i=1}^{n-1} b_i S_i. }

On the other hand, the disk contribution is
\eqn\diskk{ W_{eff}|_{\rm disk} = \CF_1.}
Combining \atlast\ with \sucfs, we find that the effective
superpotential of the theory is (up to terms with two derivatives
of $\CF_0$ and irrelevant ${\cal O}(1/\Lambda_0)$ terms)
 \eqn\gesh{ \eqalign{
 W_{eff} = & -\half \sum_{i=1}^n N_i \int_{\rb_i} y(z)dz -
 \half\sum_{I=1}^L \int_{\tilde \q_I}^{\tilde \Lambda_0}
 y(z)dz \cr
 & + \half (2N-L)W(\Lambda_0)+\half\sum_{I=1}^L W(z_I)
  - \pi i (2N-L)S + 2\pi i \tau_0 S  + 2\pi i\sum_{i=1}^{n-1}b_i S_i ,}}
where
 \eqn\tauz{2\pi i \tau_0 = \log\left({B_L\Lambda^{2N-N_f }\over
 \Lambda_0^{2N-L}}\right).}
This can be more conveniently written in terms of $R(z) =
\half(W'(z)- y(z))$ as follows:
\eqn\intR{\eqalign{ W_{eff} = & \sum_{i=1}^n N_i
\int_{\rb_i}R(z)dz + \sum_{I=1}^L \int_{\tilde
\q_I}^{\tilde\Lambda_0} R(z)dz \cr & +
(N-L)W(\Lambda_0)+\sum_{I=1}^L W(z_I) -\pi i (2N-L)S + 2\pi i
\tau_0 S + 2\pi i \sum_{i=1}^{n-1}b_i S_i. }}
This superpotential is finite in the limit when $\Lambda_0$ and
$\tilde \Lambda_0$ are taken to $P$ and $\tilde P$ respectively.
All physical information should be computed in that limit. The
logarithmic divergences in the contour integrals in \gesh\ are
cancelled by the $\Lambda_0$ dependence of $\tau_0$. Now we can
identify the finite dimensionful parameter $\Lambda $ with the
dynamical scale of the theory. Various special cases of this have
been discussed in
\refs{\CachazoJY,\NaculichHR,\OokouchiBE,\SeibergJQ}. The term
with the integer constants $b_i$ was introduced in \CachazoZK\ and
will be crucial below.  The last two terms in \gesh\ can be
written as $\sum_i \oint_{A_i} y(z) dz \int_{\B_i} ( \bar T_0
d\bar z+T_0 dz)$ where $\bar T_0 d\bar z + T_0 dz$ (which we
sometimes abbreviate simply as $T_0$) is a real differential whose
$A_i$ and $\B_i$ periods are $N_i$ and $-\tau_0 - b_i$
respectively and which also has period minus one around $\tilde
\q_I$. We will soon show that on shell $T$ and $T_0$ have the same
periods.

Here it is important to notice that, unlike cases studied in the
literature, $\tau_0$ defined by \tauz\ is not equal to the field
theory bare coupling $\tau_0^{ft}$. In field theory, the one loop
contribution to the superpotential implies that
\eqn\onll{ 2\pi i\tau_0^{ft} = (2N-N_f)\log \left(
{\Lambda_{ft}\over \Lambda_0 } \right), }
where $\Lambda_{ft}$ is the scale of the theory. Up to now, we
have made the identification $\Lambda = \Lambda_{ft}$. Recall that
$\Lambda$ was defined in the normalization of the matrix integral
\mmin. If we relax this identification, a new term has to be added
to the superpotential \suef,
\eqn\resuef{W_{eff}|_{\rm sphere} = \sum_{i=1}^n N_i {\del
\CF_0\over \del S_i} + (2N-N_f)\log \left( \Lambda_{ft}\over
\Lambda \right) S + 2\pi i\sum_{i=1}^{n-1} b_i S_i. }
Note that $W_{eff}$ in \resuef\ is independent of $\Lambda$.
Therefore, we can take $\Lambda = \Lambda_0$ and write,
\eqn\usual{W_{eff}|_{\rm sphere} = \sum_{i=1}^n N_i {\del
\CF_0\over \del S_i} + 2\pi i \tau_0^{ft} S + 2\pi
i\sum_{i=1}^{n-1} b_i S_i. }

\bigskip\noindent{\it Classical Limit Of $W_{eff}$}

A consistency check on \intR\ is to take the classical limit. In
this limit the Riemann surface $\R$ degenerates to $y^2 =
W'(z)^2$. Therefore, $R(z)\to 0$ on the first sheet and $R(z)\to
W'(z)$ on the second. In the pseudo-confining vacua, i.e., when
all singularities of $T$ are on the second sheet, the first term
in \intR\ gives $- N W(\Lambda_0) + \sum_{i=1}^n N_i W(a_i)$ while
the second gives $L W(\Lambda_0) - \sum_{I=1}^{L} W(z_I)$.
Combining this we get $W_{eff}|_{cl} = \sum_{i=1}^n N_i W(a_i)$.
On the other hand, Higgs vacua are obtained by moving some of the
poles of $T$ at points ${\tilde \q}_I$ from the second to the
first sheet. This has to be done through one of the cuts. Let us
denote by $k_i$ the number of points on the first sheet obtained
by moving the ${\tilde \q}_I$ through the $i^{th}$ cut. In this
case the contribution of the first term in \intR\ is unchanged but
from the second we get,
\eqn\higgv{ \sum_{I=1}^L r_I \int_{\q_I}^{{\tilde \Lambda}_0}
R(z)dz + \sum_{I=1}^L (1-r_I) \int_{{\tilde \q}_I}^{{\tilde
\Lambda}_0} R(z)dz = L W(\Lambda_0 ) - \!\sum_{i=1}^n k_i W(a_i) -
\!\sum_{I=1}^L (1-r_I) W(z_I). }
Recall that $r_I$ is zero if the pole of $T$ at $z=z_I$ is on the
second sheet and one if it is on the first sheet.

Using this in \intR\ we get,
\eqn\calH{ W_{eff}|_{cl} = \sum_{i=1}^n (N_i -k_i )W(a_i) +
\sum_{I=1}^L r_I W(z_I). }

\bigskip\noindent{\it Field Equations Of $W_{eff}$}

We now proceed to the study of the field equations of the
effective superpotential \gesh. Following \CachazoPR, we replace
the dynamical variables $S_i$ with the coefficients $f_i$ in
$f(z)=\sum_{i=1}^n f_i z^{i-1}$. Clearly ${\partial \over
\partial f_i} y dz= {z^{i-1} \over 2y} dz$ are holomorphic
differentials for $i=1,...,n-1$. Linear combinations of them with
coefficients which depend on $f_i$ are the holomorphic
differentials $\h_k$. In addition to those we also have $g_n
{\partial \over \partial f_n} y dz= {g_n z^{n-1} \over 2y} dz$.
Where $g_n$ was introduced so that the differential has residue
one and minus one at the poles of order one at $\tilde P$ and $P$
respectively. Again, choosing appropriate linear combinations with
the holomorphic differentials we get $\t_{P,\tilde P}$. Now
consider the equation of motion of $f_i$ by varying \gesh\ with
respect to $f_i$.  Since $W_{eff}$ is written as periods of $ydz$,
the variation is expressed in terms of periods of the holomorphic
differentials ${z^{i-1} \over 2y} dz$ and the meromorphic
differential ${z^{n-1}\over 2y}$. Appropriate linear combinations
of these give the following $n$ equations,
 \eqn\geshe{\eqalign{ 0 & = \sum_{i=1}^n N_i \int_{\B_i}\h_k +
 \sum_{I=1}^{L}\int_{{\tilde \q}_I}^{\tilde P}\h_k  + b_k \cr
 0 & =\sum_{i=1}^n N_i \int_{\rb_i}\t_{P,\tilde P} +
 \sum_{I=1}^{L}\int_{{\tilde \q}_I}^{\tilde \Lambda_0}\t_{P,\tilde P} -
 \log{\Lambda_0^{2N-L} \over B_L \Lambda^{2N-N_f}}- \pi i (2N-L).}}
We used that $\sum_{i=1}^n \oint_{A_i} \h_k=0$,
$\oint_{A_i}\t_{P,\tilde P} =0$ for $i=1,\ldots ,n-1$ and
$\oint_{A_i}\h_k =\delta_{ik}$.

Comparing with \gfink\ and \fing\ we see that these equations of
motion are equivalent to
\eqn\moti{ \tp\oint_{\b_i} T(z)dz = - b_i \qquad
\int_{\rb_n}T(z)dz =-2\pi i\tau_0 = \log{\Lambda_0^{2N-L} \over
B_L \Lambda^{2N-N_f}}.}
Here $b_i$'s are $n-1$ integers in the superpotential \gesh.
Comparing \moti\ with \diver\ we find that $\Lambda^{2N-N_f}$ in
\moti\ differs by a factor of $2^{-N_f/2}$ from $\Lambda^{2N-N_f}$
in \diver. This trivial factor can be reabsorbed in the definition
of the normalization of the matrix integral \mmin. However, we
have chosen not to include it in order to avoid cluttering the
equations.

\def\tilde{\widetilde}

\subsec{Alternative Proof}
\bigskip\noindent{\it Preliminary Discussion}

In this subsection,  we will present an alternative version of the
proof that on-shell $T\equiv T(z)dz$ and $T_0\equiv T_0(z)dz$ are
in the same cohomology class and in particular $T$ has integer
periods. To make the ideas clear, we first present a simplified
version in which we ignore the fact that some of the differential
forms of interest have poles at infinity and also at the zeroes of
the mass function $m(z)$. Then we incorporate the poles, first
those at infinity and finally those due to $m(z)$.

So to begin with, we work on a closed Riemann surface $\Sigma$ of
genus $g$, with  a set of $A$-cycles $A_i$ and $B$-cycles $B_i$,
$i=1,\dots,g$.  For any closed one-forms $\alpha $ and $\beta$ on
$\Sigma$, we define
\eqn\tinop{I(\alpha,\beta)=\int_\Sigma\alpha\wedge \beta,} which
we will usually write in the alternative form
\eqn\inop{I(\alpha,\beta)=\sum_{i=1}^g
\left(\oint_{A_i}\alpha\oint_{B_i}\beta
-\oint_{A_i}\beta\oint_{B_i}\alpha\right).} As is clear from
\tinop, $I(\alpha,\beta)=0$ if $\alpha$ and $\beta$ are both
holomorphic differentials (in that case, locally $\alpha=f(z)dz,
\,\beta=g(z)dz$, and $\alpha\wedge\beta=0$).  There is a partial
converse of this statement:

{\it (A)} If $\alpha$ is such that $I(\alpha,\beta)=0$ for any
holomorphic differential $\beta$, then $\alpha$ is in the
cohomology class of some holomorphic differential $\alpha'$, and
in particular $\alpha$ and $\alpha'$ have the same periods.

The $A$-periods $\oint_{A_i}\alpha$ of a holomorphic one-form can
be specified arbitrarily.  But once this is done, the
corresponding $B$-periods $\oint_{B_i}\alpha$ are uniquely
determined. The formula expressing the $B$-periods in terms of the
$A$-periods involves the ``period matrix'' of $\Sigma$, and
depends on the complex moduli of $\Sigma$.  In our problem, even
off-shell, that is, before we fix the complex moduli of $\Sigma$,
the $A$-periods of $T$ are equal to integers $N_i$ that depend on
the choice of a symmetry-breaking pattern.  We have the following
statement even off-shell:

{\it (B)} For any choice of moduli of $\Sigma$, the differential
$T$ is uniquely determined by being a holomorphic differential
with $\tp\oint_{A_i}T=N_i$.

If we combine statements {\it (A)} and {\it (B)}, we arrive at the
following:

{\it (C)} The cohomology class of a closed one-form $T$ is
uniquely determined by the assertions that {\it (i)}
$I(T,\omega)=0$ for any holomorphic one-form $\omega$; {\it (ii)}
$\tp\oint_{A_i}T=N_i$. If, therefore, $T_0$ is any differential
form that obeys the same conditions, then $T$ and $T_0$ have the
same periods.

In the simplified problem that we will consider first, we take the
superpotential to be \eqn\inourpro{W=\sum_iN_i\oint_{B_i}\gamma +
\sum_i b_i \oint_{A_i}\gamma,} where $\gamma$ is the meromorphic
differential $\gamma=R(z)\,dz$, while $N_i$ and $b_i$ are
integers. We can write \eqn\tinor{W=\tp I(T_0,\gamma),} where
$T_0$ is a closed (not necessarily holomorphic) one-form with
\eqn\pingo{\tp\oint_{A_i}T_0=N_i,~~~\tp\oint_{B_i}T_0=-b_i.} We
also assume that  $\gamma$ has the special property that, with
$f_i$ being the moduli of $\Sigma$, \eqn\trusgo{{\partial
\gamma\over\partial f_i}=\omega_i,} with $\omega_i$ being a basis
of the space of holomorphic differentials on $\Sigma$. (In our
actual problem, this is almost true, except that one of the
$\omega_i$ has a pole at infinity. As noted above, we will
postpone incorporating the poles until we have explained the
structure of the argument.)

Using \tinor\ and \trusgo, the condition for a critical point of
$W$, which is $\partial W/\partial f_i=0$, becomes
$I(T_0,\omega_i)=0$. But given our assumption that the $\omega_i$
are a basis of the space of holomorphic differentials on $\Sigma$,
this condition says exactly that $I(T_0,\omega)=0$ for any
holomorphic differential $\omega$. In other words, on-shell $T_0$
obeys condition {\it (i)} in statement {\it (C)}. Since $T_0$ also
obeys condition {\it (ii)} by virtue of its definition \pingo, it
follows as stated in {\it (C)} that $T$ and $T_0$ have the same
cohomology class on-shell. In particular, on-shell $\tp\oint_{B_i}
T=-b_i$.

\bigskip\noindent{\it Incorporation Of Poles At Infinity}

Our goal now, as promised above, is to describe an argument with
the same structure, reaching the same conclusion, but properly
incorporating the poles that are present in some of the
differential forms in our actual problem.

In particular, in our problem, the differential form
$\gamma=R(z)\,dz$ has a pole of order $n$ at the points $P$ and
$\tilde P$ that correspond to $z=\infty$ on the two sheets of
$\Sigma$.  The singularities of $\omega_i=\partial \gamma/\partial
f_i$ are milder; the $\omega_i$ have at worst simple poles at $P$
and $\tilde P$.

Likewise, $T$ has simple poles at $P$ and $\tilde P$. If we
include fundamental matter, then $T$ has additional poles, as we
have seen in section 4; for the moment, we consider the problem
without fundamental matter.

 The existence of all of these poles
means that the argument we have presented above does not quite fit
our physical problem. It also means that some regularization is
required in the definition of the superpotential.  We are going to
pick a regularization which is such that the proof of the on-shell
behavior of $T$ will go through directly for the regularized
theory, without having to take the regulator to infinity.

Our method of regularization in this section will be to pick
points $\Lambda_0 $ and $\tilde \Lambda_0$ that are ``near'' $P$
and $\tilde P$  and to place the poles of $T$ at $\Lambda_0$ and
$\tilde\Lambda_0$ rather than at $P$ and $\tilde P$.   We take
$\Lambda_0$ and $\tilde \Lambda_0$ to lie above the same (large)
value of $z$ on the two different sheets; later we will keep fixed
this value while varying the moduli of the hyperelliptic curve
$y^2=W'(z)^2+f(z)$. It is important to note that in this
regularization, $T$ is a solution to the anomaly equations up to
terms of order $1/\Lambda_0$. By contrast, in the regulatization
of section 5, $T$ was a solution to the anomaly equations before
taking the cut off to infinity.

Now let us explain the analogs of some assertions that we made in
our preliminary discussion.  Let $\Sigma'$ be the Riemann surface
with punctures obtained by deleting $\Lambda_0$ and
$\tilde\Lambda_0$ from the closed Riemann surface $\Sigma$.  Let
$\alpha$ be a closed one-form on $\Sigma'$. Then if $A_\Lambda$ is
a small contour circling once around $\Lambda_0$ in the
counterclockwise direction, we may have
$\oint_{A_\Lambda}\alpha\not=0$.  Likewise, let $\Sigma''$ be
obtained by deleting from $\Sigma$ the punctures $P$ and $\tilde
P$, and let $\beta$ be a closed one-form on $\Sigma''$. Then if
$A_P$ is a small contour circling once around $P$ in the
counterclockwise direction, we may have $\oint_{A_P}\beta\not=0$.
We also let $B_\Lambda$ be a path from $\tilde\Lambda_0$ to
$\Lambda_0$ (avoiding $P$ and $\tilde P$), and $B_P$ a path from
$\tilde P$ to $P$ (avoiding $\Lambda_0$ and $\tilde \Lambda_0$).
$\alpha$ has a well-defined integral over $B_P$, and $\beta$ has a
well-defined integral over $B_\Lambda$, but not vice-versa.  We
also let $A_i$ and $B_i$, $i=1,\dots, g$ be a complete set of {\it
compact} $A$- and $B$-cycles on the genus $g$ surface $\Sigma$,
chosen to avoid the points $P,\tilde P,\Lambda_0,$ and
$\tilde\Lambda_0$. In the problem studied in this paper, $g=n-1$.
We make the above choices so that $B_P$ and $B_\Lambda$ have
vanishing intersection numbers with the compact cycles. This
condition identifies $B_P$ with $B_n$ and $B_\Lambda$ with the
regularized version of $\B_n$, i.e., $\rb_n$, as defined in
section 5.

Now we define \eqn\betterdef{I(\alpha,\beta)=
\oint_{A_\Lambda}\alpha\int_{B_\Lambda}\beta -\oint_{A_P}\beta
\int_{B_P}\alpha+\sum_{i=1}^g\left(\oint_{A_i}\alpha\oint_{B_i}\beta
-\oint_{A_i}\beta\oint_{B_i}\alpha\right).} The notation $A_P,$ $
A_\Lambda$ and also $B_P$, $B_\Lambda$ is meant to suggest that
one can fruitfully think of $A_P$, $A_\Lambda$ as extra $A$-cycles
(for $\beta$ and $\alpha$, respectively), and $B_P$, $B_\Lambda$
as extra $B$-cycles.\foot{Indeed, if one glues $P$ to $\tilde P$
and $\Lambda_0 $ to $\tilde \Lambda_0$, then after smoothing the
resulting singularities, one gets a smooth surface $\widehat
\Sigma$ of genus $g+2$, with $A_P$ and $A_\Lambda$ as additional
$A$-cycles and $B_P$, $B_\Lambda$ as additional $B$-cycles.}

 Finally, by an allowed holomorphic
differential on $\Sigma''$, we will mean a holomorphic one-form on
$\Sigma''$ that has at worst simple poles at the points $P$ and
$\tilde P$. Similarly, an allowed holomorphic differential on
$\Sigma'$ has at worst simple poles at $\Lambda_0$ and
$\tilde\Lambda_0$.\foot{An allowed holomorphic differential is
precisely one that corresponds to the limit of an ordinary
holomorphic differential  on the genus $g+2$ surface $\widehat
\Sigma$ of the last footnote.}

The Riemann bilinear relations (see eqn. A.9)
 says that if $\alpha$ and $\beta$ are allowed
holomorphic differentials on $\Sigma'$ and $\Sigma''$,
respectively, then $I(\alpha,\beta)=0$.\foot{This assertion can be
deduced from the fact that $I(\alpha,\beta)=0$ for ordinary
holomorphic differentials $\alpha,\beta$ on $\widehat \Sigma$,
where here as $\widehat\Sigma $ is smooth and $\alpha$ and $\beta$
are assumed to have no poles on $\widehat \Sigma$, we use the
original definition \tinop\ of $I(\alpha,\beta)$.} Moreover, just
as in the absence of poles, this statement has a partial converse:

{\it (A$'$)} If $\alpha$ is such that $I(\alpha,\beta)=0$ for any
allowed holomorphic differential $\beta$, then $\alpha$ is in the
same cohomology class (and has the same integral over $B_P$) as
some allowed holomorphic differential $\alpha'$, and in particular
$\alpha$ and $\alpha'$ have the same periods.

The $A$-periods $\oint_{A_i}\alpha$ of a holomorphic one-form
$\alpha$ on $\Sigma'$  can be specified arbitrarily, along with
the value of $\oint_{A_\Lambda}\alpha$. But once this is done, the
corresponding $B$-periods $\oint_{B_i}\alpha$ are uniquely
determined, along with $\int_{B_P}\alpha$.  The formula expressing
the $B$-periods in terms of the $A$-periods involves a generalized
``period matrix'' of $\Sigma$, and depends on the complex moduli
of $\Sigma$.  In our problem, even off-shell, that is, before we
fix the complex moduli of $\Sigma$, the $A$-periods of $T$ are
equal to integers $N_i$ that depend on the choice of a
symmetry-breaking pattern, and we require in addition that
$\tp\oint_{A_\Lambda}T=N$ (this being the regularized version of
the assertion that $T$ has a pole at infinity with residue $N$).
We have the following statement even off-shell:

{\it (B$\; '$)} For any choice of moduli of $\Sigma$, the
differential $T$ is uniquely determined by being an allowed
holomorphic differential on $\Sigma'$ with $\tp\oint_{A_i}T=N_i$,
$\tp\oint_{A_\Lambda}T=N$.

If we combine statements {\it (A$'$)} and {\it (B$\;'$)}, we
arrive at the following:

{\it (C$\; '$)} The cohomology class of a closed one-form $T$ on
$\Sigma'$ is uniquely determined by the assertions that {\it (i)}
$I(T,\omega)=0$ for any allowed holomorphic one-form $\omega$ on
$\Sigma''$; {\it (ii)} $\tp\oint_{A_i}T=N_i$ and
$\tp\oint_{A_\Lambda}T=N$. If, therefore, $T_0$ is any
differential form that obeys the same conditions, then $T$ and
$T_0$ have the same periods.

Now, we let $T_0$ be a closed one-form on $\Sigma'$ with
\eqn\defto{ \eqalign{\tp\oint_{A_i}T_0& = N_i \cr
                     \tp\oint_{A_\Lambda}T_0& = N\cr
                     \tp\oint_{B_i}T_0& = -b_i\cr
                     \tp\int_{B_P}T_0& = -\tau_0.\cr}}
Here $\tau_0$ is the bare coupling of the gauge theory.

We define the regularized superpotential as \eqn\efto{W=\tp
I(T_0,\gamma),} where $\gamma$ is the familiar differential
$\gamma=R(z)\,dz$.  $\gamma$ is holomorphic on $\Sigma''$, but it
is not what we have called an allowed holomorphic differential, as
it has poles at $P$ and $\tilde P$ of order greater than one.
Nevertheless, the expression $I(T_0,\gamma)$ is completely
well-defined; for its definition, it suffices that $T_0$ be
regular on $\Sigma'$ and $\gamma$ on $\Sigma''$, regardless of the
nature of the behavior near the respective punctures.

Just as before, if $f_i$ are the complex moduli of $\Sigma$, then
the equation for a critical point of $W$ is $I(T_0,\omega_i)=0$
where $\omega_i=\partial\gamma/\partial f_i$.  Unlike $\gamma$,
the $\omega_i$ have only simple poles at the punctures and so are
allowed holomorphic differentials; conversely, any allowed
holomorphic differential on $\Sigma''$ is a linear combination of
the $\omega_i$.  So the condition for a critical point of $W$ is
precisely that $T_0$ obeys condition {\it (i)} in statement {\it
(C$\; '$)}.  Since $T_0$ was also defined (in \defto) to obey
condition {\it (ii)}, it follows from statement {\it (C$\; '$)}
that on-shell, in other words  when the moduli of $\Sigma$ are
chosen to get a critical point of $W$, $T_0$ and $T$ have the same
cohomology class. In particular, on-shell, $\tp\oint_{B_i}T=-b_i$
and the periods of $T$ on compact cycles are all integers.

\bigskip\noindent{\it Inclusion Of Fundamental Matter}

We developed in section 4 the necessary facts for including
fundamental matter in this discussion. Including fundamental
matter means that $T$ has simple poles, with residue 1, at certain
points $U_I$ on $\Sigma$.  (The $U_I$ may lie on the first or
second sheet; this choice will not affect our analysis and will
not be built into the notation.) $\gamma$ does not have additional
poles. So in defining $\Sigma'$, we omit the $U_I$ as well as
$\Lambda_0$ and $\tilde \Lambda_0$, while in defining $\Sigma''$,
we omit only the original punctures $P$ and $\tilde P$.  An
allowed holomorphic differential on $\Sigma'$ or $\Sigma''$ has
only simple poles at the punctures.

We let $A_I$ be a small contour circling $U_I$ once in the
counterclockwise direction, and $B_I$ a path from $U_I$ to $P$
avoiding all other punctures, and with vanishing intersection with
all the other cycles. For $\alpha$ a closed one-form on $\Sigma'$
and $\beta$ a closed one-form on $\Sigma''$, we define
\eqn\weddef{I(\alpha,\beta)=
\oint_{A_\Lambda}\alpha\int_{B_\Lambda}\beta
+\sum_I\oint_{A_I}\alpha\int_{B_I}\beta-\oint_{A_P}\beta
\int_{B_P}\alpha+
\sum_{i=1}^g\left(\oint_{A_i}\alpha\oint_{B_i}\beta
-\oint_{A_i}\beta\oint_{B_i}\alpha\right).}

All previous statements have obvious analogs.  If $\alpha$ and
$\beta$ are, respectively, allowed holomorphic differentials on
$\Sigma'$ and $\Sigma''$, then $I(\alpha,\beta)=0$.  This
statement has the partial converse stated as {\it (A$'$)} above.
We will omit copying this statement.

Moreover, {\it (B$\; '$)} and {\it (C$\; '$)} have obvious
analogs. The $A$-periods $\oint_{A_i}\alpha$ of a holomorphic
one-form $\alpha$ on $\Sigma'$ along with the values of
$\oint_{A_\Lambda}\alpha$ and $\oint_{A_I}\alpha$
 can be specified arbitrarily. Once this is done, the
corresponding $B$-periods $\oint_{B_i}\alpha$ are uniquely
determined along with $\int_{B_P}\alpha$, and depend on the moduli
of $\Sigma$.  However:

{\it (B$\; ''$)} For any choice of moduli of $\Sigma$, the
differential $T$ is uniquely determined by being an allowed
holomorphic differential on $\Sigma'$ with $\tp\oint_{A_i}T=N_i$,
$\tp\oint_{A_\Lambda}T=N$, $\tp\oint_{A_I}T=1$.

If we combine statements {\it (A$'$)} and {\it (B$\; ''$)}, we
arrive at the following:

{\it (C$\; ''$)} The cohomology class of a closed one-form $T$ on
$\Sigma'$ is uniquely determined by the assertions that {\it (i)}
$I(T,\omega)=0$ for any allowed holomorphic one-form $\omega$ on
$\Sigma''$; {\it (ii)} $\tp\oint_{A_i}T=N_i$,
$\tp\oint_{A_\Lambda}T=N$, and $\tp\oint_{A_I}T=1$. If, therefore,
$T_0$ is any differential form that obeys the same conditions,
then $T$ and $T_0$ have the same periods.

The rest of the argument is hopefully clear.  We  let $T_0$ be a
closed one-form on $\Sigma'$ with \eqn\defto{ \eqalign{
 \tp\oint_{A_i}T_0& = N_i \cr
                     \tp\oint_{A_\Lambda}T_0& = N\cr
                     \tp\oint_{A_I}T_0&=1
                     \cr
                     \tp\oint_{B_i}T_0& = - b_i\cr
                     \tp\int_{B_P}T_0& = -\tau_0.\cr}}
We define the superpotential as $W=\tp I(T_0,\gamma)$.  The
condition for a critical point is, exactly as before, that
$I(T_0,\omega)=0$ for every allowed holomorphic differential
$\omega$ on $\Sigma''$. Hence on-shell, $T_0$ obeys the conditions
in {\it (C$\; ''$)}. Therefore, $T$ and $T_0$ have the same
periods and in particular $\tp\oint_{B_i}T=-b_i$.

The superpotential $W_{eff} = \tp I(T_0,\gamma)$ has been
characterized so far by conditions on its critical points. This
still allows for an additive piece independent of the complex
structure of the Riemann surface $\R$. In order to fix this piece,
consider the classical limit in which the Riemann surface
degenerates to $y^2 = W'(z)^2$. This leads to $R(z)=0$ on the
first sheet and $R(z)=W'(z)$ on the second. Recalling that $\gamma
= R(z)dz$ and the conditions
\defto, we get
\eqn\ccopi{ \left. \tp I(T_0,\gamma )\right|_{cl} = \sum_{i=1}^n
N_i W(a_i)+ (L-N)W(\Lambda_0 ) - \sum_{I=1}^L W(z_I). }
On the other hand the classical superpotential should be
$W_{eff}|_{cl} = \sum_{i=1}^n N_i W(a_i)$. Therefore,
\eqn\suwah{ W_{eff} = \tp I(T_0,\gamma) + (N-L)W(\Lambda_0) +
\sum_{I=1}^L W(z_I). }
In order to compare this superpotential with \intR, we should
recall that the regularization procedure is different. The only
place this is relevant is in the integral of $T$ along $B_P$ and
the $B_I$'s. $T$ has a pole at $\Lambda_0$ and $\tilde\Lambda_0$.
However, the $T$ solution to the anomaly equations has a pole at
$P$ and $\tilde P$. The difference of the integrals can be shown
to be equal to half the period of $T$ around $A_{\Lambda}$ up to
terms of order $1/\Lambda_0$. This implies that the difference is
equal to $\pi i$. This accounts for the $\pi i (L-2N) S$ term in
\intR.

\newsec{Interpolating Between Different Vacua}

We start this discussion by considering a certain process
involving variation of the parameters in our model.  We consider
an example that can be understood semiclassically.

Let as usual $a_j, \,j=1,\dots,n$ be the zeroes of $W'(z)$, and
$z_I$ the zeroes of the mass function $B(z)=\det \,m(z)$.  We
consider a classical vacuum with $N_j$ eigenvalues of $\Phi$ set
equal to $a_j$, leaving a $U(N_j)$ factor in the low energy gauge
group. Now consider a process in which, keeping the other
parameters fixed, one of the $z_I$ circles counterclockwise once
around one particular $a_i$ in the complex plane, chosen so that
$N_i\not= 0$.  Since $z_I$ determines one of the quark mass
parameters in the effective $U(N_i)$ theory, in this process one
of those parameters changes in phase by $2\pi$.  For example, if
$N_f=1$ and $m(z)=z_I-z$, then the quark mass parameter is
$z_I-a_i$, and changes in phase by $2\pi$ when $z_I$ circles
around $a_i$. In the process, the theta angle $\theta_i$ of the
$U(N_i)$ theory increases by $2\pi$.

\ifig\oneloopdiag{Interpolation between different $\theta_i$
values. $(a)$ Original configuration. Cycle $\b_i$ is shown. On
the upper sheet it is oriented from the $i^{th}$ cut to the
$n^{th}$ cut while on the lower sheet from the $n^{th}$ cut to the
$i^{th}$ cut. The points $\q_I$ (black dot) and $\tilde \q_I$
(white dot) are enclosed by the $C_I$ and $\tilde C_I$ contours
respectively. The dashed line represents the motion in the lower
sheet of $\tilde\q_I$ as it circles once around the $i^{th}$ cut.
$q_I$ also moves for $z(q_I)=z(\tilde q_I)$. $(b)$ Final
configuration after the points $\tilde \q_I$ and $q_I$ return to
their original position completing the trajectory around the
$i^{th}$ cut. Also shown is the deformation of the original $\b_i$
cycle. Clearly, the new $\b_i$ cycle is given by $\b_i -\tilde C_I
+ C_I$.} {\epsfxsize=0.7\hsize\epsfbox{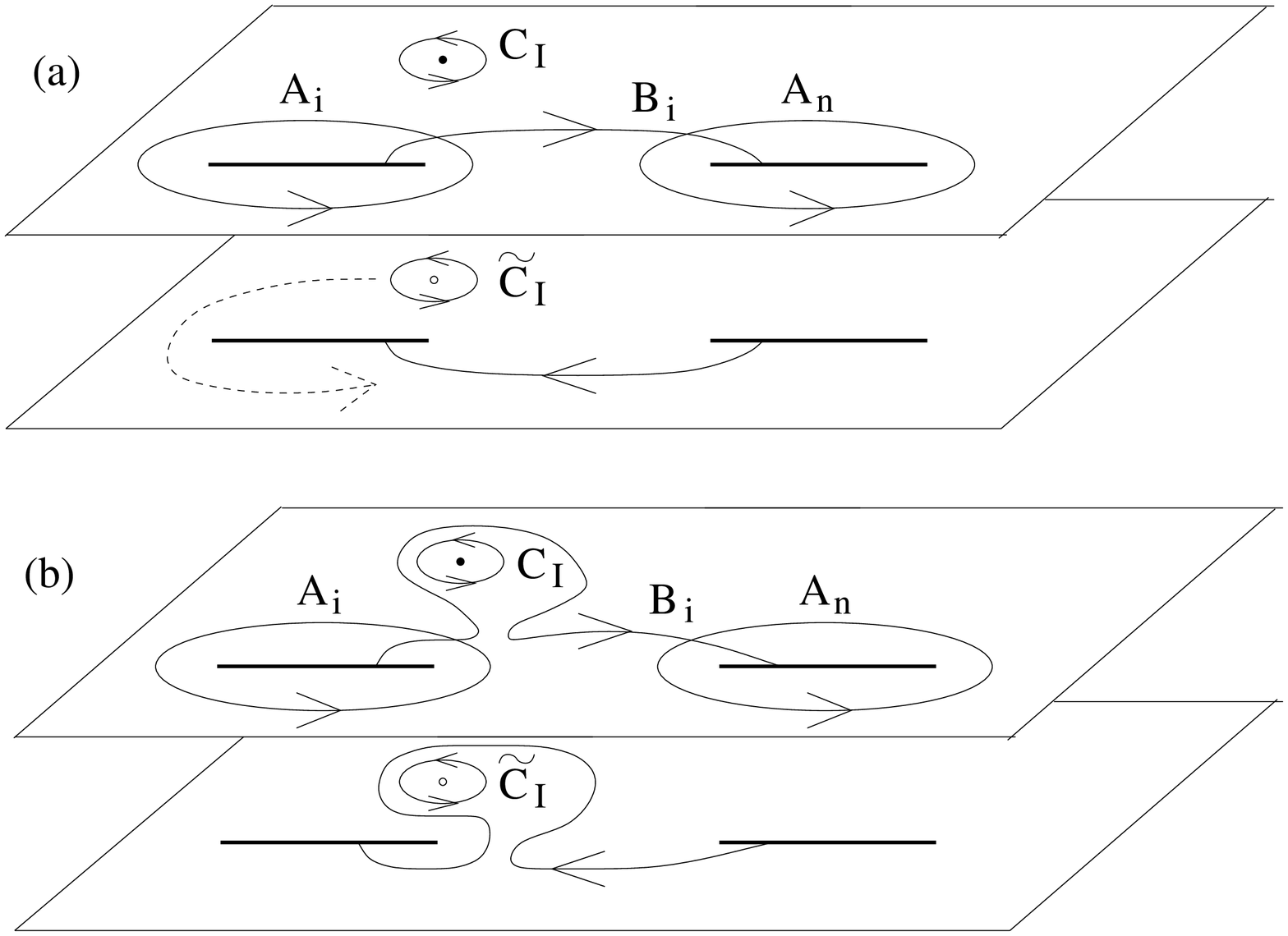}}

The periods of $T(z)\,dz$ integrated over certain cycles measure
differences between theta angles \CachazoZK.  In particular, if
the $\b_i$ are the compact cycles described in figure 2, and we
set \eqn\tomigo{b_i=-{\tp}\oint_{\b_i}T(z)\,dz,} then we have
 \eqn\omigo{\theta_i-\theta_n=2\pi b_i.}
When $z_I$ circles around $a_i$ and $\theta_i\to \theta_i+2\pi$,
we must therefore have \eqn\eqres{b_i\to b_i+1.}

We could also have obtained the last formula by an argument
similar to that used in sections 2 and 4 to explain the change in
the $N_i$ when $z_I$ passes through a cut.  We will carry out the
discussion first in the pseudo-confining phase with $r_I=0$. For
any given $z_I$, the differential $T(z)\,dz$ has a pole of residue
1 at the point $\tilde \q_I$ corresponding to $z_I$ on the second
sheet. When $z_I$ migrates around $a_i$, $\tilde \q_I$ sweeps out
a closed path $\tilde Y$ that intersects the cycle $\b_i$. To
avoid the singularity, one should move the cycle $\b_i$
continuously in \tomigo\ as $z_I$ is moved.  By the time $z_I$
returns to its starting point, $\b_i$ returns as $\b_i-\tilde
C_I$, where $\tilde C_I$ is a small contour that wraps once around
$\tilde \q_I$. Since $\tp \oint_{C_I} T(z)\,dz=1$, this leads to
\eqres.

What happens if we consider the same motion of $z_I$, but this
time in the Higgs phase, in which the pole associated with $z_I$
is on the first sheet at $\q_I$ (rather than on the second sheet
at $\tilde \q_I$)?  In this case, as  $z_I$ migrates around $a_i$,
$\q_I$ sweeps out a closed path $Y$ which intersects the contour
$\b_i$, but now, since the part of the contour $\b_I$ that is on
the first sheet runs in the opposite direction (if projected to
the $z$-plane) from that in the second sheet, the sign of the
intersection number is reversed, \eqn\jugvc{Y\cap \b_i=-\tilde
Y\cap \b_i.} To avoid the singularity, the cycle $\b_i$ must again
be varied continuously; it returns as $\b_i+C_I$, where $C_I$ is a
small contour surrounding $\q_I$. We get $\b_i+C_I$, compared to
$\b_i-\tilde C_I$ in the last paragraph, because of the relative
minus sign in \jugvc.  The opposite sign implies that in the Higgs
phase, with $r_I=1$, the shift in $b_i$ when $z_I$ migrates around
$a_i$ is $b_i\to b_i-1$, compared to $b_i\to b_i+1$ in the last
paragraph. The sign reversal can be readily explained in the field
theory. For $r_I=0$, as $z_I\to a_i$, components of $Q$ and
$\tilde Q$ comprising a chiral multiplet in the fundamental
representation of $U(N_i)$ (plus its conjugate) become massless;
their contribution to the $U(N_i)$ beta function is $+1$.  For
$r_I=1$, as $z_I\to a_i$, the expectation value of the adjoint
superfield $\Phi$ as well as $Q$ and $\tilde Q$ change in such a
way (see \Phiexh) that the unbroken $U(N_i)$ is extended to an
unbroken $U(N_i+1)$.  The extra fields that become massless
consist of a vector multiplet in the fundamental representation
(the gauge fields that extend $U(N_i)$ to $U(N_i+1)$) along with a
chiral multiplet (coming from $Q$) plus similar multiplets in the
conjugate representation.  The total contribution to the beta
function of these fields is $-3+ 1=-2$, where $-3$ is the
contribution of the vector multiplet. The opposite sign beta
function determines the opposite sign contribution to $b_i$ from
integrating out these multiplets near $a_i\approx z_I$.  In
getting the monodromy of the theta angle, the 2 in the beta
function is cancelled by the fact that, according to \qdet, the
expectation value of $Q$ (and hence the masses produced by the
Higgs mechanism) are proportional to $\sqrt{z_I-a_i}$. More
explicitly, the instanton factor of the low energy gauge group
$\Lambda_L^p$ depends on the beta function.  Near $a_i\approx z_I$
it scales like $\Lambda_L^p \sim z_I-a_i$ for $r_I=0$ and
$\Lambda_L^p \sim (z_I-a_i)^{-1}$ for $r_I=1$.  Therefore
depending on whether $r_I=0$ or $r_I=1$, $b_i$ changes by opposite
amounts when $z_I$ winds around $a_i$.

We can express the result \eqres\ in the following fashion.  While
$z_I$ is circling around $a_i$, the position $\tilde \q_I$ of the
pole in $T$ associated with $z_I$ migrates around a certain closed
path $C$ on our Riemann surface $\Sigma$. $C$ has intersection
number 1 with the closed contour $\b_i$, as indicated in the
figure 3. The transformation law of $b_i$ is $b_i\to b_i+\b_i\cap
C$, where $\b_i\cap C$ is the intersection number of $\b_i$ and
$C$. We can state this more invariantly by saying that for any
closed contour $D$ in $\Sigma$, when a pole of $T$ passes around a
contour $C$, the period of $T$ changes by
\eqn\yumigo{\tp\oint_DT(z)\,dz\rightarrow \tp\oint_DT(z)\,dz+D\cap
C.} In fact, of the $A_j$ and $\b_j$ cycles in figure 2, only
$\b_i$ has a nonzero intersection number with $C$; and only the
period $b_i=-\tp\oint_{\b_i}T(z)\,dz$ is expected to change in a
semiclassical process in which $z_I$ circles around one of the
$a_i$.  An even more invariant way to say this is that the
cohomology class of $\tp T(z) \,dz$ changes in such a process by
$\tp T(z)\,dz\to \tp T(z)\,dz+[C]$, where $[C]$ is the cohomology
class that is Poincar\'e dual to $C$.

In section 4, we considered a process in which one of the $z_I$
(or more precisely the corresponding $\q_I$) passes through the
cut $A_i$ onto the second sheet. We showed that in this process,
there is a transition from a pseudo-confining phase to a Higgs
phase, and a transformation $N_i\to N_i-1$.  Consider now the
following process that starts and ends at the same phase, which we
can take to be pseudo-confining:  we let $\q_I$ start on the first
sheet, pass through the cut $A_i$ to the second sheet, and then
return to the first sheet by passing through the cut $A_{i'}$.
Clearly, in this process the $N$'s transform by $N_i\to N_i-1$,
$N_{i'}\to N_{i'}+1$, with no change in the others. This can be
regarded as another illustration of eqn. \yumigo.  In this case,
one of the poles of $T$ has made a closed circuit $C$ from the
second sheet to the first and back again; of the canonical $A_j$
and $\b_j$ cycles of figure 2, the ones that have non-zero
intersection with $C$ are $A_i$ and $A_{i'}$.

Classically, the pseudo-confining phases are labeled by the
integers $N_i$ with the restriction $\sum_iN_i=N$, as well as by
the $b_i$.
 By iterating the processes that we have so far described,
we can (with a caveat explained shortly) change in an arbitrary
fashion the $N_i$ and $b_i$, so these operations connect all of
the pseudo-confining phases.  By combining these operations with a
further process in which some $\q_I$ move to the second sheet and
stay there, we can similarly connect pseudo-confining phases to
arbitrary Higgs phases.  The ability to connect all the phases in
this way should not come as a complete surprise, because in this
model, in contrast to that studied in \CachazoZK, the presence of
fields in the fundamental representations prevents the existence
of precise order parameters involving Wilson and 't Hooft loops of
the underlying $U(N)$ gauge theory.

\bigskip\noindent{\it On-Shell And Off-Shell Continuation}

This discussion is subject to an important caveat.  As in section
4, we have here imagined changing $m(z)$ while keeping fixed the
Riemann surface $\Sigma$.  But $\Sigma$ depends on the resolvent
$R(z)$ and thus on the gluino condensates $S_i$; on-shell, the
$S_i$ should really be determined by extremizing the effective
superpotential $W_{eff}$, and $W_{eff}$ depends on $m(z)$.  So in
general, when $m(z)$ is changed, the $S_i$ and $\Sigma$ will
change. A more delicate question now arises: if while changing
$m(z)$, one adjusts the $S_i$ so as to remain on-shell, is the
continuation between the different phases still possible?

The process described  involving a change in the $b_i$ when the
$z_I$ circle around the $a_i$ can be seen semiclassically and so
can certainly occur on-shell. The process in which $\q_I$ passes
through a cut, changing some $N_i$, is more delicate as the region
with $\q_I$ near the cut is strongly coupled.  Intuitively, we
believe that when the $N_i$ are large enough, the back-reaction
due to a single eigenvalue passing through the cut is small, and
hence it should be possible for all of these processes to occur
on-shell.  A bit of evidence in this direction is that for ${\cal
N}=2$ super Yang-Mills theory with $SU(2)$ gauge group and one
flavor, perturbed slightly to ${\cal N}=1$ by a mass for the
adjoint superfield, it is possible to have a smooth transition
from a branch with a condensed monopole and ``confinement'' to a
branch with a condensed quark and a ``Higgs effect'' \SeibergAJ.
The model is a special case (with $N=2$, $N_f=1$, linear $m(z)$,
and quadratic $W$) of the model considered here;  the
interpolation found in \SeibergAJ\ should correspond in our
current language to a process in which $\q_I$ passes through a
cut. (Moreover, the restriction to linear $m(z)$, etc., should be
inessential to the behavior when $\q_I$ is near a cut.) So this
indicates that a process lowering $N_i$ by 1 by passing an
eigenvalue through a cut can occur even when $N_i=2$. Subsequent
work on ${\cal N}=2$ theories with matter \tyu\ similarly
indicates that such processes can occur for all $N_i>2$.

For $N_i=1$, however, we cannot expect to see a process in which
yet another $\q_I$ passes through a cut, reducing $N_i$ to zero.
In such a process, the number of unbroken $U(1)$ factors in the
low energy gauge group would change; this cannot occur
continuously. It must be that the back reaction of a change in
$m(z)$ on $\Sigma$ is crucial when a zero of $m(z)$ approaches a
cut $A_i$ with $N_i=1$.  This is not surprising, since the
infrared behavior of $U(1)$ gauge theory with massless matter is
completely different from that of $U(1)$ gauge theory without
massless matter, so the effect of the zero of $m(z)$ on the cut
will be large.

\newsec{The Theory With $W(\Phi)=0$}

As an application of the general results of the previous sections,
we now study the special case of the theory with $W(\Phi)=0$. This
theory has a moduli space of vacua $\CM$ parametrized by
$\Phi_{cl}$ modulo conjugation by a unitary matrix. Alternatively,
$\CM$ can be parametrized by the expectation values
 \eqn\gaui{u_k={1\over k} \langle \Tr \Phi^k \rangle, \qquad\qquad
 k=1,...,N .}
Our goal is to compute the various observables as a function of
$\Phi_{cl}$ or $u_k$.

We start by examining the structure of $\CM$ in the classical
theory.  At a generic point in $\CM$, the fundamental $U(N)$ gauge
symmetry is broken to $U(1)^N$ and all the quarks are massive.  At
special points, there are singularities.   If $k$ of the
eigenvalues of $\Phi_{cl}$ are equal, there is in addition an
unbroken $SU(k)$ symmetry.  When an eigenvalues of $\Phi_{cl}$
coincides with one of the zeros of ${\hat B}(z)$, say $z_I$, there
is also a massless ``electron.''

In the quantum theory, we expect the $SU(k)$ singularities to be
replaced by monopole points, but the points with massless
electrons are IR free and are essentially unchanged. This
semiclassical picture is identical to that in the $U(N)$ $\CN=2$
theory with $L$ flavors.  Therefore, we expect the moduli space to
be described by the hyperelliptic curve $ y_0^2=\det (z -
\Phi_{cl})^2-4\Lambda^{2N-N_f}{\hat B}(z)$.  We will soon see that
this is indeed the case.  It is therefore clear that if $L>2N$ the
metric on $\CM$ is not positive definite at large $\Phi$ and the
theory is not sensible.  Note that this can happen even for
$N_f<2N$ where the gauge coupling is asymptotically free.  For
arbitrary polynomials $m_f^{\tilde f}$, the standard link between
the beta function and the behavior of $\tau$ on the moduli space
is not present.

As in \CachazoPR, in order to study the theory at a generic point
in $\CM$, we deform it with a generic superpotential $W(\Phi)$ of
degree $N+1$. We parametrize it as
 \eqn\wppa{W'(z)=g_N\prod_{i=1}^N(z-a_i).}
We study the theory for small $g_N$ and then set $g_N$ to zero. We
consider the vacuum in which the $U(N)$ gauge symmetry is broken
to $U(1)^N$, i.e. in which all $N_i=1$. This leads to
 \eqn\eqhh{\eqalign{
 y^2= & \; W'(z)^2+f(z)= g_N^2 y_0^2 \cr
 y_0^2= & \; P^2(z)-4\Lambda^{2N-N_f}{\hat B}(z), \qquad\qquad P(z)=
 \det (z - \Phi_{cl})}}
Equation \eqhh\ follows from \sigval\ since $Q(z)$ and $H(z)$ are
constants in this case.

For $L=\deg({\hat B}) \le N-1$ this equation easily determines
$W'(z)=g_N P(z)$ and $f(z)= -4g_N\Lambda^{2N-N_f} {\hat B}(z)$.
Starting with a tree level superpotential with parameters $a_i$ in
\wppa, the vacuum of the system is at a point in $\CM$ which is
labeled by $\Phi_{cl}={\rm diag}(a_1,a_2,...,a_N)$. When we
examine $T(z)$ below, we will see that for $k=1,...,N$ in this
case $u_k={1\over k} \Tr \Phi_{cl}^k $.  For $N\le L < 2N$ this
identification is not so easy. The eigenvalues of $\Phi_{cl}$ are
not the same as $a_i$ in \wppa\ and $u_k={1\over k} \langle \Tr
\Phi^k \rangle \not= \Tr \Phi_{cl}^k$.

It is clear that $R(z)$ is proportional to $g_N$ and therefore it
vanishes when $g_N \to 0$.  Similarly, $M(z)$ is proportional to
$g_N$ and it vanishes in this limit.  However,
 \eqn\tsolw{T(z) = {\partial \over \partial z} \log \left( P(z) +
 y_0(z)\right) = {P'(z) \over y_0(z)} +{{\hat B}'(z) \over 2{\hat B}(z)} - {P(z){\hat B}'(z)
 \over 2y_0(z){\hat B}(z)}}
has a smooth nonzero limit as $g_N\to 0$, and therefore \tsolw\ is
valid also when $W=0$.

As a check, from \eqhh\ we learn that $y_0(z_I)=P(z_I)$ (recall
that we use the value in the first sheet), and since $P(z)$ is a
polynomial,
 \eqn\spec{\sum_{I=1}^L {y_0(z_I)\over 2 (z-z_I)}=\sum_{I=1}^L {P(z_I)\over
 2(z-z_I)} =\left[{P(z){\hat B}'(z) \over 2{\hat B}(z)}\right]_-}
Using this in \anoeqa\ equation \Tsolv\ becomes
 \eqn\Tsolva{ T(z)={ {\hat B}'(z)\over
 2{\hat B}(z)}-{P(z){\hat B}'(z)\over 2y_0(z) {\hat B}(z)}+ {\tilde c(z) \over y_0(z)}}
with another polynomial $\tilde c(z)$.  It agrees with our general
solution \tsolw\ for ${\tilde c}(z) = P'(z)$.

Now that we know $T(z)$, we can calculate
 \eqn\uktz{u_k={1 \over k} \langle \Tr \Phi^k \rangle = {1 \over 2\pi i k}
 \oint dz z^k T(z).}
For $L<N$, we easily find that the gauge invariant coordinates on
$\CM$ are given by $u_k={1 \over k} \Tr \Phi_{cl}^k$
($k=1,...,N$). The observables $u_k$ with $k>N$ are not given by
their classical values ${1 \over k} \Tr \Phi_{cl}^k$, but receive
quantum corrections which are polynomials in the instanton factor
$\Lambda^{2N-N_f}$.  These are quantum deformations of the
classical relations in the chiral ring.  This generalizes the
result of \CachazoRY\ in the pure gauge theory to the case with
matter.

For $N\le L < 2N$, we find a more dramatic result.  Even the
coordinates on $\CM$, which are the generators of the chiral ring,
are not given by the classical expressions but receive instanton
corrections $u_k={1 \over k} \Tr \Phi_{cl}^k +
\CO(\Lambda^{2N-N_f})$ ($k=1,...,N$).

As we said above, these results are the same as in the $\CN=2$
theory with $L$ flavors.  Equation \tsolw\ is thus an expression
for $T(z)$ in the $\CN=2$ theory with matter.

Now that we have explored the theory with $W(\Phi)=0$, we can turn
on an arbitrary superpotential $W(\Phi)$.  The effect of this
superpotential can be analyzed as a small perturbation.  From the
structure of our equations is it clear that it cannot change the
value of $T$ and that $R$ and $M$ are linear in $g_n$.  The main
effect of the superpotential is to choose vacua -- choose
$\Phi_{cl}$ in $\CM$.  Then various monopoles and electrons at
that point condense and lift some of the photons in these vacua.
But the curve $y_0^2=P(z)^2-4\Lambda^{2N-N_f}{\hat B}(z)$ in these
vacua is unchanged.  The theory with nonzero $W$ inherits the
curve from the theory with vanishing $W$.

\bigskip
\centerline{\bf Acknowledgements}

It is a pleasure to thank M. Douglas and J. Maldacena for helpful
discussions. This work was supported in part by DOE grant
\#DE-FG02-90ER40542 and NSF grant PHY-0070928 to IAS.

\appendix{A}{Riemann Bilinear Relations.}

In this appendix we will give a brief review of the main tools
from the theory of Riemann surfaces used in this work. Most of the
results presented here are standard and are given for the reader's
convenience. For this review we followed \fark.

A Riemann surface of genus $g$ can be thought of as a $4g$ polygon
with some identifications of its edges. This representation is
particularly useful to get relations among the integral of
meromorphic one-forms along various cycles of the Riemann surface.
These are known as Riemann bilinear relations. Although, this
requires a particular choice of cycles and therefore breaking of
modular invariance explicitly, the relations obtained this way can
then be interpreted in any other basis of cycles by properly
deforming the contours. This is the source of the integer
ambiguity in all the formulas that refer to it in the main text of
this paper. In sections 3 and 5 where the different cycles are
explicit, the same choice has been made, which is consistent with
the discussion in this appendix.

For our purposes it is enough to consider meromorphic one-forms
with at most simple poles. Let $\H$ have simple poles at $T$ and
$Q$ and $\G$ to have simple poles at $R$ and $S$. The Riemann
surface $\R$ of genus $g$ on which $\H$ is defined has puncture at
$T$ and $Q$. In order to get a simply connected representation of
it, we introduce cuts as shown in figure 4, the new surface $\R'$
is simply connected and $\H$ can be written as $df$ on it.

Let us compute $\int_{\del \R'} f \G $, where $\del \R'$ is the
boundary of $\R'$ shown in figure 4. We can evaluate this in two
different ways. The first is to deform the contour to enclose the
poles of $\G$ which we assume not to be located on $\del \R'$.
Using Cauchy's formula this leads to,
\eqn\onew{\int_{\del \R'} f \G  = 2\pi i f(R){\rm res}_R \G + 2\pi
i f(S){\rm res}_S \G .}

\ifig\oneloopdiag{Polygon $\R'$ representing a Riemann surface
$\R$ of genus $g=3$. Cuts from $O$ to $T$ and to $Q$ are
introduced in order to make $\R'$ simply connected. The path $C$
runs from $O$ to $T$ and back around the cut $OT$ and from $O$ to
$Q$ and back around the cut $OQ$. Also shown is the reference
point $z_0$ in the definition of $f(z)$. The points $z$ and $z'$
are identified. Dashed lines represent the contours used in
integrals below.} {\epsfxsize=0.6\hsize\epsfbox{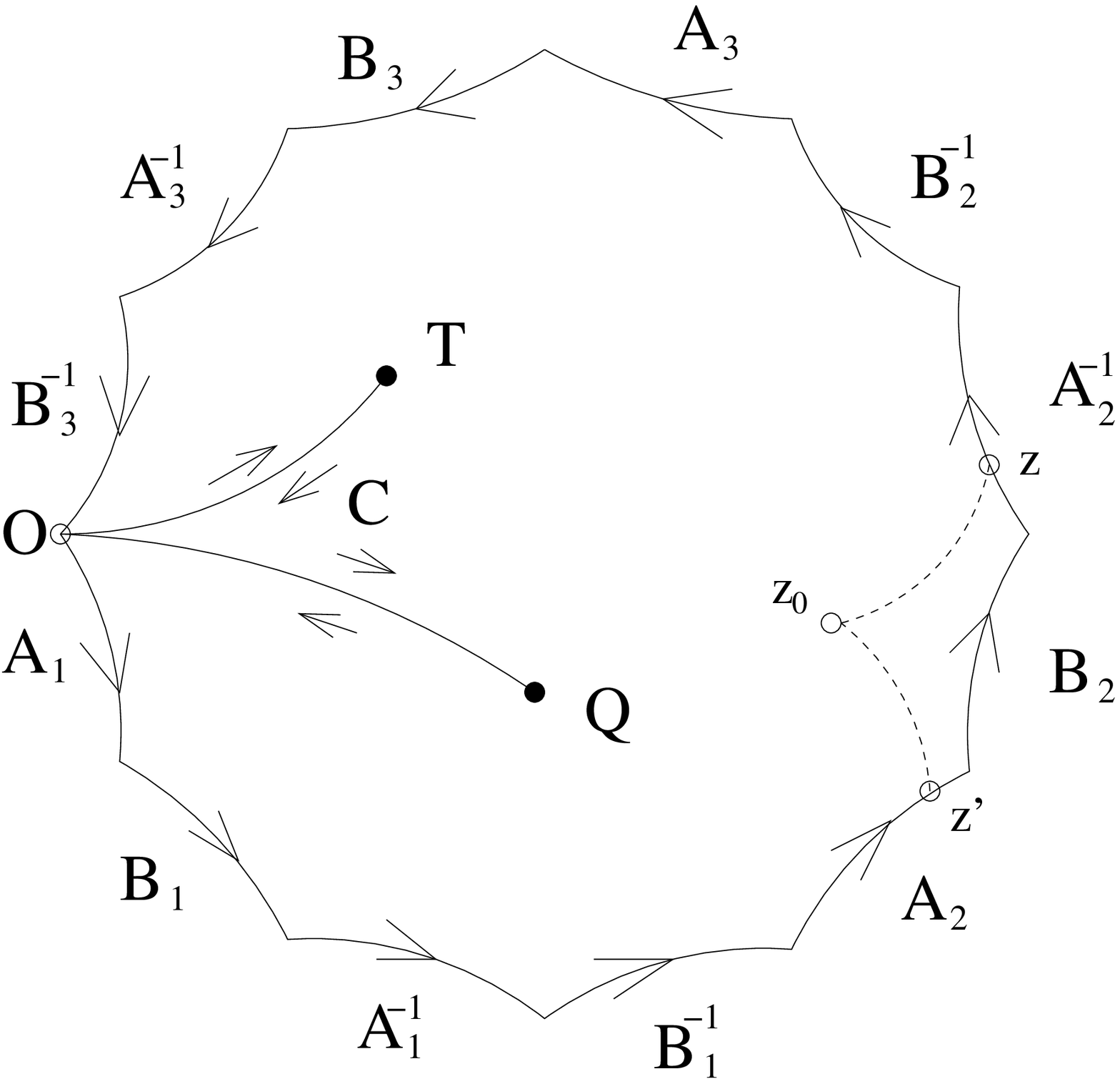}}

Since $\G$ does not have any more poles, ${\rm res}_R \G = - {\rm
res}_S \G$.  Using $\H=df$ we can write \onew\ as,
\eqn\onefin{\int_{\del \R'} f \G =  2\pi i\left( {\rm res}_R \G
\int_S^R \H \right). }
The second way is to use the contour in figure 4. From there we
conclude that,
\eqn\lkqw{ \int_{\del \R'} f \G = \sum_{j=1}^g \left( \int_{A_j}f
\G +\int_{\b_j}f \G +\int_{A^{-1}_j}f \G + \int_{\b^{-1}_j}f \G
\right) + \int_{C}f \G }
Letting $f(z) = \int_{z_0}^z \H$ where $z_0$ is an arbitrary point
in $\R'$ (see figure 4), we can compute,
\eqn\whips{ \int_{A_j}f \G + \int_{A^{-1}_j}f \G =
\int_{A_j}\left( \int_{z_0}^z \H - \int_{z_0}^{z'}\H  \right) \G }
where $z$ and $z'$ are points on $A_j$ and $A^{-1}_j$ respectively
that are identified (see figure 4). The two integrals of $\H$ can
be deformed to a single integral over the $\b_j$ cycle to give,
\eqn\parres{ \int_{A_j}f \G + \int_{A^{-1}_j}f \G = -\int_{\b_j}
\H \int_{A_j} \G .}
Likewise,
\eqn\lwsw{\int_{\b_j}f \G + \int_{\b^{-1}_j}f \G = \int_{\b_j} \G
\int_{A_j} \H .}
The last integral in \lkqw\ runs over the contour $C$ and can be
computed by realizing that the value of $f$ on one side of the cut
$OT$ differs from the value on the other side by $\oint_T \H$.
Similarly, it differs by $\oint_Q \H$ between the two sides of the
cut $OQ$. This implies that,
\eqn\kpqw{ \int_{C}f \G = \oint_T\H \int_O^T\G + \oint_Q\H
\int_O^Q \G  =  {\rm res}_T\;\H \int_Q^T\G .}
Combining \lkqw, \parres, \lwsw\ and \kpqw\ we have,
\eqn\sonw{ \int_{\del \R'} f \G = \sum_{i=1}^g \left(
\oint_{A_i}\H \oint_{\b_i}\G - \oint_{A_i}\G \oint_{\b_i}\H
\right) + 2\pi i{\rm res}_T \H \int_Q^T \G.}
Here we have also assumed that $\H$ does not have any more poles.

Comparing \onefin\ and \sonw\ we get the Riemann bilinear relation
which all relations used in this work are special cases of,
\eqn\gere{ 2\pi i {\rm res}_R \G \int_S^R \H - 2\pi i {\rm res}_T
\H \int_Q^T \G
 = \sum_{i=1}^g \left( \oint_{A_i}\H \oint_{\b_i}\G -
\oint_{A_i}\G\oint_{\b_i}\H \right). }
Before illustrating the use of this formula, let us point out that
the cycles used to define the edges of the 2g-polygon are not the
same as the cycles in the double sheeted representation of the
same genus g hyperelliptic Riemann surface. These two sets of
cycles are in general related by an $Sp(2g,{\bf Z})$
transformation. We have not distinguished them since the right
hand side of \gere\ is clearly $Sp(2g, {\bf Z})$ invariant. A more
important distinction arises when the two sets of basis differ by
the orientation of the resulting surfaces. In this case \gere\ can
be used by reversing the orientation of the $B_i$ cycles. This
will be important in the first application of \gere\ given next.

As a first illustration of the use of \gere\ let us derive \billl.
In this case we have $\H = \t_{p_1,p_2}$ and $\G=\h_k$. Recall
that $\t_{p_1,p_2}$ has zero periods around the $A$ cycles and
poles of order one with residue $-1$ and $1$ at $p_1$ and $p_2$
respectively. On the other hand $\h_k$ is a holomorphic one form
with $\oint_{A_i}\h_k = \delta_{ik}$. The choice of orientation in
figure 2 is the opposite to that of figure 4. Therefore, as
discussed before, the sign of integrals over $B$ cycles should be
reversed. Combining this in \gere\ we get \billl,
\eqn\debill{ \tp\oint_{\b_k}\t_{p_1,\p_2}  = \int_{p_2}^{p_1}\h_k.
}

As a second illustration, let us derive \riebi. Take $\H =
\t_{T,Q}$ and $\G=\t_{R,S}$. Recalling that these differentials
have zero periods around the $A$ cycles, the right hand side of
\gere\ is zero. This leads to,
\eqn\rbed{ {\rm res}_R\t_{R,S}  \int_S^R \t_{T,Q}  = {\rm res}_T
\t_{T,Q} \int_Q^T \t_{R,S} .}
Using that ${\rm res}_R\t_{R,S}=-1$ and ${\rm res}_T \t_{T,Q}=-1$
we get \riebi.

The derivation of \gere\ and therefore of \riebi\ assumes that all
points $R,S,T,Q$ are distinct. In section 3 we need a degenerate
version of this, i.e., the case when $T=R$. The integral in \rbed\
are divergent in this case and need a regulator. Without loss of
generality we can choose local coordinates vanishing at $T$. Let
us then take $R$ at $z=\epsilon$. For $\epsilon \neq 0$ we can use
\rbed,
\eqn\rezt{\int_S^\epsilon \t_{0,Q}  =  \int_Q^0 \t_{\epsilon ,S}.
}
We need to express the integral on the right hand side in terms of
$\t_{0,S}$ and a regularized contour. The two integrals are not
the same but are related by exchanging the end of the contour with
the location of the pole. This operation introduces the integral
around half a contour around the pole and therefore gives a half.
Let us check this explicitly. We need to compute,
\eqn\fpq{\int_Q^0 \t_{\epsilon ,S} - \int_Q^\epsilon \t_{0 ,S}. }
In the local coordinates we chose, $\t_{\epsilon ,S}=- {1\over
z-\epsilon}dz + {\rm regular}$ and $\t_{0,S} =-{1\over z}dz+ {\rm
regular}$. The regular terms will not contribute to \fpq\ in the
limit $\epsilon \to 0$, and therefore will not be considered.
Explicit evaluation of \fpq\ gives $-\log (-\epsilon )+ \log
\epsilon+ \CO(\epsilon) =-\pi i + \CO(\epsilon) $. Therefore,
\eqn\finwer{\int_Q^0 \t_{\epsilon ,S}= \int_Q^\epsilon \t_{0 ,S}
-\pi i + \CO(\epsilon)\ . }
Using this in \rezt\ we get the general form of \useex,
\eqn\geseex{ \int_S^\epsilon \t_{0,Q}  =\int_Q^\epsilon \t_{0 ,S}
-\pi i + \CO(\epsilon)\ . }

\appendix{B}{Sphere Contribution To Effective Superpotential}

In this appendix we will carry out the computation of ${\del
\CF_0\over \del S_i} $ from the saddle point approximation result
\fzer. The answer is used to compute the sphere contribution to
the effective superpotential $W_{eff}$ of the field theory \suef.
We will use standard matrix model techniques to calculate $\CF_0$
(for a review see e.g.\ \DiFrancescoNW), and will express the
answer in terms of contour integrals. Similar computations can be
found elsewhere (e.g.\ in \NaculichHR), but we will keep terms
that were not important for the analysis there, and were not
explicitly shown.  We will also manipulate only finite quantities
making all dependence on the cutoff $\Lambda_0$ explicit.

Recall that our normalization for the eigenvalue density is $\int
d\lambda \rho (\lambda )=S$. Therefore, the filling fractions are
\eqn\fillf{ S_j = \int_{a_j^-}^{a_j^+}d\lambda \rho(\lambda ) .}
Consider \fzer\ modified to include chemical potentials $\mu_i$.
\eqn\che{ \CF_0(\mu_i, \rho ) = \int d\lambda
\rho(\lambda)W(\lambda ) -\int\!\!\int d\lambda d\lambda'
\rho(\lambda )\rho(\lambda' )\log\left|{\lambda-\lambda'\over
\Lambda }\right| - \sum_{i=1}^n\mu_i \int_{a_i^-}^{a_i^+}d\lambda
\rho(\lambda ). }
This is a functional of $\rho(\lambda)$ and a function of
$\mu_i$'s. $S_i$'s will be introduced via a Legendre transform.

The variation of \che\ with respect to $\rho(\lambda)$ with
$\lambda$ a point in the $i$-th cut is
\eqn\vafun{ {\delta \CF_0\over \delta \rho(\lambda )} = W(\lambda)
- 2 \int d\lambda' \rho(\lambda')\log\left|{\lambda-\lambda'\over
\Lambda }\right| - \mu_i  =  0 .}
This is the equation of motion of $\rho(\lambda)$.  It follows
from it that
 \eqn\derf{ W'(\lambda ) - 2 \; \CP\!\!\!\int d\lambda' {\rho
 (\lambda' )\over \lambda - \lambda'} =0}
where $\CP$ denotes the principal part.

Although we do not need it here, we add for completeness a
derivation of the loop equation.  By multiplying \derf\ by
$\rho(\lambda)\over z- \lambda $ and integrating over $\lambda$ we
derive
 \eqn\loopeqaa{\eqalign{
 \int d\lambda {\rho(\lambda) W'(\lambda )\over z- \lambda} =&
 2 \; \CP\!\!\!\int d\lambda d\lambda'{  \rho(\lambda
 )\rho(\lambda') \over(z- \lambda )( \lambda - \lambda')}\cr
 =&\CP\!\!\!\int d\lambda d\lambda' {\rho(\lambda )\rho(\lambda')
 \over \lambda - \lambda'} \left({1\over z-  \lambda}-{1\over
 z- \lambda' }\right)\cr
 =& \int d\lambda d\lambda' {\rho(\lambda )\rho(\lambda')
 \over (z- \lambda )( z- \lambda')}\cr
 =&\left(\int d\lambda {\rho(\lambda ) \over z- \lambda }\right)^2
 }}
Since $\rho(\lambda)=-{1 \over 2\pi i}{\rm disc}R(\lambda)$, $\int
d\lambda \rho(\lambda ) F(\lambda)= \sum_i \tp\oint_{A_i} dw R(w)
F(w)$ for any function $F(w)$ which is analytic on the cuts $A_i$.
A special case of this is $\int d\lambda {\rho(\lambda) \over z-
\lambda } = \sum_i \tp\oint_{A_i}dw {R(w)\over z- w}  = -
\tp\oint_z dw {R(w)\over z-w} =
 R(z)$.  Using $\int d\lambda {\rho(\lambda) W'(\lambda) \over
z-\lambda} = \left[ W'(z)\int d\lambda {\rho(\lambda) \over z-
\lambda} \right]_-=\left[ W'(z) R(z)\right]_-$, we derive from
\loopeqaa\ the loop equation $\left[ W'(z) R(z)\right]_-=R(z)^2$.

{}From the equation of motion \vafun\ we solve for $\langle
\rho(\lambda )\rangle$ as a function of $\mu_1,...,\mu_n$.  This
leads to the free energy with fixed chemical potentials
\eqn\newf{\CF_0(\mu_1,...,\mu_n) = \CF_0(\mu_1,...,\mu_n ,\langle
\rho(\lambda )\rangle ) .}
Under a Legendre transformation
\eqn\letr{ \CF_0 (S_1,...,S_n) = \CF_0(\mu_1,...,\mu_n)+
\sum_{i=1}^n \mu_i S_i}
where
\eqn\conj{ {\del \CF_0(\mu_1,...,\mu_n)\over \del \mu_i} + S_i = 0
}
is used to write $\mu_i = \mu_i (S_1,...,S_n)$.  The Legendre
transform (equation of motion of $S_i$) determines
\eqn\pard{ {\del \CF_0(S_1,...,S_n)\over \del S_i} = \mu_i. }
Using \vafun\ we find
\eqn\vapsr{ {\del \CF_0(S_1,...,S_n)\over \del S_i} = W(\lambda) -
2 \int d\lambda' \rho(\lambda')\log\left|{\lambda-\lambda'\over
\Lambda }\right|. }
Recall that $\lambda$ is a point in the $i$-th cut. However, the
combination appearing in \vapsr\ is independent of $\lambda$ (see
\derf). Therefore, a convenient $\lambda$ can be chosen to
evaluate \vapsr. Take $\lambda = a_i^+$, i.e. one of the ends of
the $i$-th cut. The integral over the cuts of $\rho(\lambda')$ can
be replaced by a sum of integrals around the $A_j$ cycles of
$R(z)$ as follows
\eqn\acti{{\del \CF_0(S_1,...,S_n)\over \del S_i} = W(a_i^+) - 2
\sum_{j=1}^n \tp\oint_{A_j} dz R(z)\log\left( {z - a_i^+ \over
\Lambda }\right). }
Note that the integrand is not an analytic function on the cut
$A_i$ but the logarithmic singularity does not lead to a
divergence there. The continuation of the function $\log | \;
a_i^+ -\lambda'|$ into a complex function might introduce more
terms which are linear combinations of $2\pi i S_j$ with integer
coefficients. These terms are irrelevant for our computation since
they correspond to a choice of the winding number of the
regularized contours $\rb_j$ around the $A_k$ cycles as will be
clear from the final expression.

We now deform the contours in \acti\ to a contour that starts at
$\Lambda_0$ winds around on a large circle of radius
$|\Lambda_0|$, moves in along the cut of the logarithm to $a_i^+$
and back to $\Lambda_0$ along the other side of this cut.  Using
$R(z) = {S\over z} + \CO(1/z^2)$, this leads to
 \eqn\defcon{\eqalign{{ \del \CF_0(S_1,...,S_n)\over \del S_i}
 =&W(a_i^+) + 2 \int_{a_i^+}^{\Lambda_0}R(z)dz - 2 S \log
 \left( - {\Lambda_0\over \Lambda }\right)+
{\cal O}(1/\Lambda_0)\cr
 =&W(\Lambda_0) - \int_{a_i^+}^{\Lambda_0}y dz -
 2 S \log \left( - {\Lambda_0\over\Lambda }\right)+
{\cal O}(1/\Lambda_0).}}
where we used $2R(z)= W'(z) - y(z)$.  Since $y(z)$ vanishes at
$a_i^+$, we can write it as
\eqn\conesp{{\del \CF_0\over \del S_i}=  -\half\int_{\rb_i}y(z)dz
+ W(\Lambda_0) - 2S \log\left(-{\Lambda_0\over \Lambda } \right)+
{\cal O}(1/\Lambda_0). }
Now that the answer is written as a contour integral, the contour
can be moved through the cut $A_i$ and does not have to pass
through $a_i^+$.

Finally, we can write the sphere contribution to the effective
superpotential from \suef,
\eqn\atlast{ W_{eff}|_{\rm sphere } = -\half \sum_{i=1}^n
N_i\int_{\rb_i}y(z)dz + 2\pi i\sum_{i=1}^{n-1} b_i S_i +  N
W(\Lambda_0) - 2N S \log\left(-{\Lambda_0\over \Lambda } \right).
}
up to terms of order $1/\Lambda_0$.

Before concluding this appendix let us note that the free energy
$\CF_0$ is a homogeneous function of degree 2 in the combination
of $S_1,...,S_n$ and $g_0,...,g_n$. In order to see this, consider
the definition,
\eqn\defG{ \exp\left( -{{\hat N}^2\over S^2}\CF_0 + ... \right) =
{1\over {\rm vol}\; U(\hat N) }\int {d \hat\Phi \over
\Lambda^{{\hat N}^2}} \exp \left( -{{\hat N}\over S} \Tr
W(\hat\Phi )\right). }
where the ellipses represent higher order terms in the $1/{\hat
N}$ expansion.

Recalling that $W(z) = \sum_{k=0}^n {g_k\over k+1}z^{k+1}$, it is
simple to see that $\CF_0$ can only depend on the couplings
through the combination $g_k / S$. In addition, $\CF_0$ can only
depend on $S_i$'s through the filling fractions $S_i/ S$. Finally,
note that from the exponential in \defG, the free energy has to be
of the form,
\eqn\scal{\CF_0 = S^2 f\left( {g_0\over S},...,{g_n\over S},
{S_1\over S}, ...,{S_n\over S} \right).}
This proves the statement that $\CF_0$ is homogeneous of degree 2
in $g_i$'s and $S_i$'s; i.e.
\eqn\degt{ 2 \CF_0 =\sum_{k=0}^n g_k {\del \CF_0 \over \del g_k} +
 \sum_{i=1}^n S_i {\del \CF_0 \over \del S_i}.}

\appendix{C}{Strong Coupling Analysis}

Once $T(z)$ has been found for the theory without superpotential,
we can repeat the strong coupling analysis of \CachazoZK\ for the
case when the deformation is by $W(\Phi)$ of degree $N+1$. This
will be a consistency check of the previous results since it leads
to the identification of the curves as in \eqhh\ and it is carried
out in the opposite limit, i.e.\ in the strong coupling regime.
The main reason this computation is nontrivial is that in the
quantum theory $W(\Phi)$ is not given by $W(\Phi_{cl})$ but we
have to be careful about the instanton corrections.

The superpotential for $\Phi$ is introduced as a small deformation
of the theory. This deformation leads to an effective
superpotential for the vev's of ${1\over k}\Tr \Phi^k$ denoted by
$U_k$.

These vev's are computed from the generating function
\eqn\ggen{ \left\langle \Tr {1\over z-\Phi }  \right\rangle = {N
\over z} +\sum_{l=1}^{\infty} {l U_l\over z^{l+1}} }
{}From \conk\ we also have
\eqn\assu{ \left\langle \Tr {1\over z-\Phi } \right\rangle =
{\partial\over \partial z}\log \left( P(z) + \sqrt{P^2(z)-4
\Lambda^{2N-N_f}\hat B(z)} \right)}

Using \ggen\ and \assu\ we get
\eqn\pp{P(z) = z^{N} \exp \left( -\sum_{i=1}^\infty {U_i\over z^i}
\right) +\Lambda^{2N-N_f}{\hat B(z) \over z^{N}} \exp \left(
\sum_{i=1}^\infty {U_i\over z^i} \right)  }
Given that $P(z)$ is a polynomial of degree $N$, \assu\ can be
thought of as determining the coefficients of $P(z)$ and $U_k$ for
$k\geq N+1$ in terms of $U_l$ for $l=1,\ldots ,N$. It is important
to note that $P(z)$ is not equal to $\langle \det (z-\Phi
)\rangle$. From \pp\ the relation can be seen to be
\eqn\rell{ P(z) = \langle \det (z-\Phi )\rangle + \left[
\Lambda^{2N-N_f}{\hat B(z) \over z^{N}} \exp \left(
\sum_{i=1}^\infty {U_i\over z^i} \right)\right]_+ }

In order to use the deformation by the superpotential $W(\Phi)$ to
fix a point in the space of $U_k$'s for $k=1,\ldots , N$, we have
to take the degree of $W$ to be $N+1$. This implies that $U_{N+1}$
has to be written in terms of the lower traces. We will consider
all $U_k$'s to be independent and impose the relation from \pp\ as
a constraint.

The deformation by the superpotential $W(\Phi)$ gives an effective
superpotential for $U_k$'s of the form
\eqn\effsup{ W = \sum_{r=0}^{N}g_r U_{r+1}  + R_0 \oint \left(
z^{N} \exp \left( -\sum_{i=1}^\infty {U_i\over z^i} \right)
+\Lambda^{2N-N_f}{\hat B(z) \over z^{N}} \exp \left(
\sum_{i=1}^\infty {U_i\over z^i} \right)   \right) dz }
where $R_0$ is a Lagrange multiplier imposing the constraint that
enforces $U_{N+1}$ to be a function of the lower $U_l$'s.

Taking variations with respect to $U_{r+1}$, we get
\eqn\vrrll{g_r = R_0 \oint \left( z^{N-r-1} \exp \left(
-\sum_{i=1}^\infty {U_i\over z^i} \right)-\Lambda^{2N-N_f}{\hat
B(z) \over z^{N+r+1}} \exp \left( \sum_{i=1}^\infty {U_i\over z^i}
\right)  \right) dz  }
Using this we can compute $W'(x) = \sum_{r=0}^{N}g_r x^r$ to be
\eqn\suyy{ W'(x) = \oint \left(  z^{N} \exp \left(
-\sum_{i=1}^\infty {U_i\over z^i} \right)-\Lambda^{2N-N_f}{\hat
B(z) \over z^{N}} \exp \left( \sum_{i=1}^\infty {U_i\over z^i}
\right) \right) {dz\over (z-x)} }
where we have assumed that $L \leq 2N$ in order to extend the sums
over $r$ from $N$ to infinity and perform the sum to get
$(z-x)^{-1}$. Here $z$ is assumed to be inside the contour of
integration.

On the constraint surface,
\eqn\conff{ z^{N} \exp \left( -\sum_{i=1}^\infty {U_i\over z^i}
\right) = {1\over 2} \left( P(z)+ \sqrt{P^2(z) -4
\Lambda^{2N-N_f}\hat B(z) } \right).}
Using this in \suyy, we get
\eqn\sumat{ W'(z) = R_0 \oint { \sqrt{P^2(z)-4
\Lambda^{2N-N_f}\hat B(z)} \over (z-x)}dz}
Letting $g_{N}=1$, this equation implies that
\eqn\conjj{ y^2 = P^2(z)-4 \Lambda^{2N-N_f}\hat B(z) = W'(z)^2 +
{\cal O}(x^{N-1}).}
This is consistent with the result found in \sigval\ and \eqhh,
i.e.\
\eqn\fgen{  y^2 = P^2(z)-4 \Lambda^{2N-N_f}\hat B(z) = W'(z)^2 +
f(z).}

\listrefs

\end